\RequirePackage{fix-cm}
\documentclass[smallextended]{svjour3}       
\smartqed  
\usepackage{graphicx}

  \usepackage{cite}

\usepackage{tikz}
\usepackage{amsmath,amstext,amsfonts,amssymb}
\usepackage{amsbsy}
\usepackage{lettrine}
\usepackage{float}
\newcommand{\ds}{\displaystyle}

\newcommand{\mc}{\mathcal}
\newcommand{\tc}{\textcolor}

\newtheorem{Proposition}{Proposition}
\begin{document}

\title{Efficient Packet Transmission in Wireless Ad Hoc Networks with Partially Informed Nodes
}


\author{Sara Berri \and Samson Lasaulce  \and Mohammed Said Radjef
}


\institute{S. Berri \at
 Research Unit LaMOS (Modeling and Optimization of Systems), Faculty of Exact Sciences, University of Bejaia, Bejaia, 06000, Algeria\\
L2S (CNRS-CentraleSupelec-Univ. Paris Saclay), Gif-sur-Yvette, France\\
              \email{sara.berri@l2s.centralesupelec.fr}             \\
            \emph{Present address: Telecom ParisTech-Univ. Paris Saclay, Paris, France}  
          \and
           S. Lasaulce \at
              L2S (CNRS-CentraleSupelec-Univ. Paris Saclay), Gif-sur-Yvette, France\\
               \email{Samson.LASAULCE@lss.supelec.fr}
               \and
             M.S. Radjef \at
            Research Unit LaMOS (Modeling and Optimization of Systems), Faculty of Exact Sciences, University of Bejaia, Bejaia, 06000, Algeria\\
             \email{radjefms@gmail.com}
}

\date{Received: date / Accepted: date}

\maketitle

\begin{abstract}
One formal way of studying cooperation and incentive mechanisms in wireless ad hoc networks is to use game theory. In this respect, simple interaction models such as the forwarder's dilemma have been proposed and used successfully. However, this type of models is not suited to account for possible fluctuations of the wireless links of the network. Additionally, it does not allow one to study the way a node transmits its own packets. At last, the repeated game models used in the related literature do not allow the important scenario of nodes with partial information (about the link state and nodes actions) to be studied. One of the contributions of the present work is precisely to provide a general approach to integrate all of these aspects. Second, the best performance the nodes can achieve under partial information is fully characterized for a general form of utilities. Third, we derive an equilibrium transmission strategy which allows a node to adapt its transmit power levels and packet forwarding rate to link fluctuations and other nodes actions. The derived results are illustrated through a detailed numerical analysis for a network model built from a generalized version of the forwarder's dilemma. The analysis shows in particular that the proposed strategy is able to operate in presence of channel fluctuations and to perform significantly better than existing transmission mechanisms (e.g., in terms of consumed network energy).

\end{abstract}

\maketitle

\section{Introduction}\label{sec:introduction}

In wireless ad hoc networks, nodes are interdependent. One node needs the assistance of neighboring nodes to relay the packets or messages it wants to send to the receiver(s). Therefore, nodes are in the situation where they have to relay packets, but have at the same time to manage the energy they spend for helping other nodes. \tc{black}{To capture the tradeoff between a cooperative behavior (which is necessary to convey information through an ad hoc network) and a selfish behavior (which is necessary to manage the node energy), the authors of \cite{hub} proposed a simple but efficient model.} Their modeling has been found to be very important and insightful in the literature of ad hoc networks, as advocated by the many papers where it is exploited. The model consists in \tc{black}{assuming, whatever the size of the network, that the local node interaction only involves two neighboring nodes having a decision-making role; one of the virtues of considering the interaction to be local is the possibility of designing distributed transmission strategies.} In the original model of \cite{hub}, a node has two possible choices namely, forward or drop the packets it receives from the neighboring node. \tc{black}{As shown in \cite{hub}, modeling the problem at hand as a game appears to be natural and relevant; in the corresponding game, the node utility function consists of the addition of a data rate term (which is maximized when the other node forwards its packets) and an energy term (which is maximized when the node does not forward the packets of the other node).} At the Nash equilibrium of the corresponding strategic-form \tc{black}{static} game (called the forwarder's dilemma in the corresponding literature), nodes don't transmit at all \tc{black}{\cite{hub}}. To avoid this, cooperation has to be stimulated e.g., by studying the repeated interaction between the nodes \cite{hub}, \cite{b1}, \cite{c2i7} or by implementing incentive mechanisms \cite{c2i7}, \cite{c2i8}, \cite{a2}, \cite{c2i9}, \cite{c2i1}, \cite{s4}, \cite{h1}. \tc{black}{The vast majority of incentive mechanisms either rely on the idea of reputation \cite{c2i7}, \cite{a2}, \cite{charles2010}  or the use of a credit system \cite{c2i8}, \cite{c2i9}, \cite{ref2016}, \cite{tvtCred}. While providing an efficient solution, all the corresponding models still have some limitations, especially regarding link quality fluctuations and partial knowledge at the nodes; indeed,  they do not take into account the quality of the link between the transmitting and the receiving nodes, which may be an important issue since the link quality may strongly fluctuate
if it is wireless. The solution in \cite{c2i1} referred to as ICARUS (for hybrId inCentive mechAnism for coopeRation stimUlation in ad hoc networkS), combines the two ideas namely, reputation and credit system, but it is not suited to scenarios where the actions of the other nodes are not perfectly observed, which results e.g., in inappropriate punishment (a node is declared selfish while it is cooperative) and therefore in a loss of efficiency. Additionally, in \cite{c2i1}, when a node is out of credit, the transmission is blocked and the node cannot send any packet anymore; this might be not practical in some wireless networks where a certain quality-of-service has to be provided. Also \cite{c2i1} proposes a mechanism to regulate the credit when a node has an excessive number of credits, but the proposed  mechanism may be too complex. At last but not least, in \cite{c2i1} no result is provided on the strategic stability property, which is important and even necessary to make the network robust against selfish deviations.} The purpose of this paper is precisely to overcome the \tc{black}{limitations of the aforementioned previous works. More precisely, the contributions of the present paper are as follows.}\\

\noindent
\tc{black}{$\blacktriangleright$ The first key technical difference with the closely related works is that the proposed formulation accounts for the possible presence of quality fluctuations of the different links that are involved in the considered local interaction model. In particular, this leads us to a game model which generalizes the existing models since the forwarding game has now a state and the discrete action sets are arbitrary, not just binary; additionally, the node does not only choose the cooperation power but also the power used to send its own packets.}\\
\tc{black}{$\blacktriangleright$ An important and useful contribution of the paper is to characterize the feasible utility region of the considered problem, by exploiting implementability theorems provided by recent works  \cite{benja}, \cite{larrousse}, \cite{larrousse-tit-2015}. This problem is known to be non-trivial in the presence of partial information and constitutes a determining element of folk theorems; this difficult problem turns out to be solvable in the proposed reasonable setting (the channel gains are i.i.d. and the observation structure is memoryless). The knowledge of
the utility region is very useful since it allows one to measure the efficiency of any distributed algorithm relying on the assumed partial information.}\\
\tc{black}{$\blacktriangleright$ A third contribution of the paper is that we provide a new transmission strategy whose main features is to be able to deal with the presence of fluctuating link qualities and to be efficient. To design the proposed strategy, we show that the derived utility region can be used in a constructive manner to obtain efficient operating points, and propose a new incentive mechanism to ensure that these points are equilibrium points. The proposed incentive mechanism combines the ideas of credit and reputation. To our knowledge, the closest existing incentive mechanism to the one proposed in the present paper is given in \cite{c2i1}. Here, we go further by dealing with the problem of imperfect observation and that of credit outage or excess. Indeed, the credit evolution law we propose in this paper prevents, by construction, the number of credits from being too large; therefore one does not need to resort to an additional credit regulation mechanism, which may be too complex.}\\
\tc{black}{$\blacktriangleright$ In addition to the above analytical contributions, we provide a numerical study which demonstrates the relevance of the proposed approach. Compared to the closest transmission strategies, significant gains are obtained both in terms of packet forwarding rate, network consumed power, and combined utilities. As a sample result, the network consumed power is shown to be divided by more than two w.r.t. state-of-the art strategies \cite{c2i1} and \cite{c2i5}.}

The remainder of the paper is organized as follows. In Sec. \ref{sec:System-model}, we present the system model; the assumed local interaction model \tc{black}{involves two neighboring} nodes of an ad hoc network \tc{black}{with arbitrary size} and generalizes the famous model introduced in \cite{hub}. The associated static game model is also provided in Sec. \ref{sec:System-model}. In Sec. \ref{sec:game-formulation}, the repeated game formulation of the generalized packet forwarding problem is provided; one salient feature of the proposed model is that partial information is assumed both for the network state and the nodes actions. In Sec. \ref{sec:region-algorithm}, the feasible utility region of the studied repeated game with partial observation is fully characterized. We also provide an algorithm to determine power control policies that are shown to be globally efficient in Sec.~\ref{sec:performance-analysis}. The proposed incentive mechanism and equilibrium transmission strategy are provided in Sec. \ref{sec:equilibrium}; the proposed transmission strategy allows both the packet forwarding rate and the transmit power to be adapted. A detailed numerical performance analysis is conducted in Sec. \ref{sec:performance-analysis}. Sec. \ref{sec7} concludes the paper.

\section{System model}
\label{sec:System-model}

\begin{figure}[h!]
\begin{center}
\includegraphics[width=8.5cm,height=6cm]{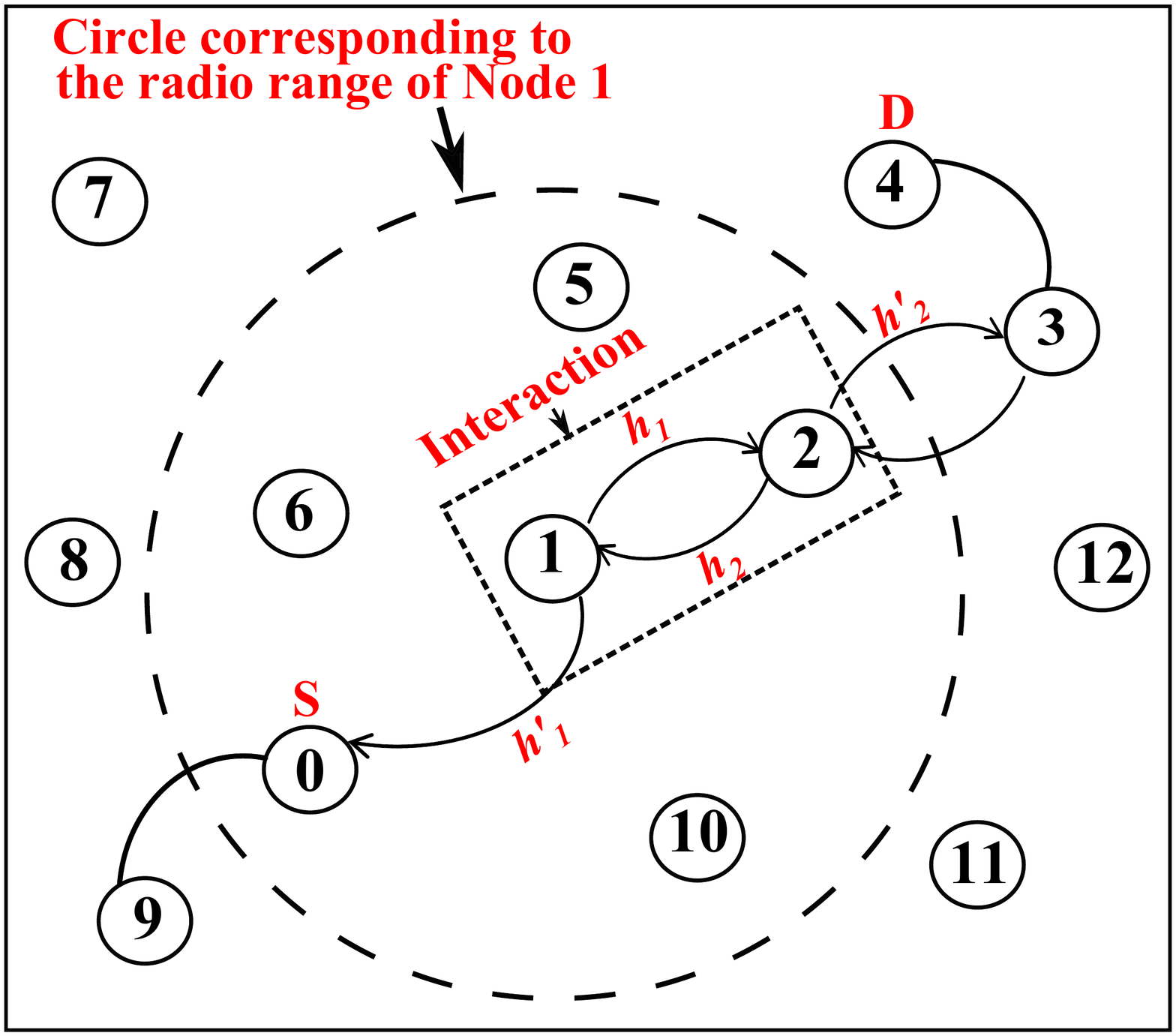}\\
\end{center}
\caption{\tc{black}{In this example, the focus is on what Node 1 does to allow Node 0 (source) to route its packets to Node 4 (destination). The dashed circle represents the radio range for Node 1 and defines its neighbors. To ensure a distributed design, two key elements are exploited: a) Node 1 adopts its transmission behavior to each of its neighbors. Here, Node 1 interacts with Node 2 (indicated by the dotted box); b) Only the available knowledge of the quality of the most influential links is accounted for (denoted generically by $h_1, h_1', h_2, h_2'$)}}\label{fig-interaction-model}
\end{figure}

The present work concerns wireless ad hoc networks namely, networks in which a source node needs the assistance of other nodes to communicate with the destination node(s). As well motivated in related papers such as \cite{c2i7}, \cite{ji-tmc-2010}, \cite{a55}, we will assume the interaction among nodes to be local i.e., it only involves neighboring nodes. This means that the network can have an arbitrary size and topology but a node only considers local interactions to take its decision although it effectively interacts with more nodes. One of the virtues of such an interaction model is to be able \textbf{to design distributed transmission strategies} for every node. More specifically, we will assume the famous model of \cite{hub} in which local interactions take place in a pairwise manner, which not only allow us to design distributed strategies but also to easily compare the proposed transmission strategy with existing strategies. The key idea of this relevant model is to take advantage of the fact that the network is wireless to simplify the interaction model. For a given node, the dominant interaction will only involve its closest neighbors (see Fig.~\ref{fig-interaction-model}). If several neighboring nodes lie within the radio range of the considered node, then it is assumed to have several pairwise interactions in parallel, as explained in detail in the numerical part.

The nodes are assumed to be non-malicious i.e., each of them does not aim at damaging the communication of the other. Additionally, they are assumed to operate in an imperfect promiscuous mode, which means that each node imperfectly overhears all packets forwarded by their neighbors. The proposed model generalizes the model \cite{hub} for at least four reasons. First, the action of a node has two components instead of one: the transmit power used to help the other node, which is denoted by $p_i'$; the transmit power used to send its own packets, which is denoted by $p_i$. Second, the transmit power levels are not assumed to be binary but to lie in a general discrete set\footnote{The notation $|.|$ stands for the cardinality of the considered set.} $\mathcal{P}_{i} =\mathcal{P}_{i}' = \mathcal{P}=$ $\{P_1,P_2, ...,P_L\} =$ $\{P_{\min},\ldots,P_{\max}\}$, $|  \mathcal{P}_{i} |=$ $|\mathcal{P}_{i}'| = $ $|\mathcal{P}|=L$. Assuming that the sets are discrete is of practical interest, as there exist wireless communication standards in which the power can only be decreased or increased by step and in which quantized wireless channel state information (CSI) is used (see e.g., \cite{david}, \cite{sesia-book-2009}). \tc{black}{Similarly, the channel may be quantized to define operating modes (e.g., MCS -modulation coding scheme) used by the transmitter. Even when the effective channel is continuous, assuming it to be discrete in the model and algorithm part may be very relevant. At last, note that from the limiting performance characterization point of view, the analysis of the continuous case follows from the discrete case but the converse is not true \cite{discret}}. As a third new feature compared to \cite{hub}, the considered model accounts for the possible fluctuations of the quality of each link. With each link a non-negative scalar is associated, which is called the \textit{channel gain} of the considered link. For a node, the channel gains of the links used to send its own packets and to help the other node, are denoted by $h_i$ and $h_i'$, respectively. These channel gains are assumed to lie in discrete sets (of states): $\mathcal{H}_{i}=\mathcal{H'}_{i}=  \mathcal{H}  =\{h_{\min},\ldots,h_{\max} \}$ with $|\mathcal{H}_{i}|=$ $|\mathcal{H'}_{i}|= $ $|\mathcal{H}| = H$; the realizations of each channel gain will be assumed to be i.i.d.. Technically, continuous channel gains might be assumed. But, as done in the information theory literature for establishing coding theorems, we address the discrete case in the first place, since the continuous case can be obtained by classical arguments (such as assuming standard probability spaces), whereas the converse is not true. Now, from the practical aspect, quantizing the channel gains typically induces a small performance loss compared to the continuous case; one figure assuming a typical scenario illustrates this. The corresponding channel gain model naturally applies to time-selective frequency flat fading single-input single-output channels. If the channel gain is interpreted as the combined effect of path loss and shadowing, our model can also be used to study more general channel models such as MIMO channels. Fourth, the utility function of a node has a more general form than in the forwarder's dilemma. The \textit{instantaneous utility function} for Node $i\in\{1,2\}$ expresses as follows:
\begin{equation}\label{eq51}
   u_{i}(a_0,a_{1},a_{2})=\varphi(\mathrm{SNR}_{i})-\alpha(p_i+p'_i),
\end{equation}
where:
\begin{itemize}
\item $a_0=(h_1,h'_1,h_2,h'_2)$ is the \textit{global channel or network state}. The corresponding set will be denoted by $\mc{A}_0=\mathcal{H}_{1} \times \mathcal{H'}_{1} \times \mathcal{H}_{2} \times \mathcal{H'}_{2} = \mathcal{H}^4$;
\item $a_i = (p_i, p_i')$ is the \textit{action} of Node $i\in\{1,2\}$;
\item the function $\varphi$ is a communication efficiency function which represents the \textit{packet success rate}. It is assumed to be increasing and lie in $[0,1]$. A typical choice for $\varphi$ is for example, $\varphi(x) = (1-e^{-x})^\ell$, $\ell$ being the number of symbols per packet (see e.g., \cite{goodman-pcom-2000}, \cite{meshkati-jsac-2006}, \cite{lasaulce-twc-2009}) or $\varphi(x) = e^{-\frac{c}{x}}$ with $c=2^r -1$, $r$ being the spectral efficiency in bit/s per Hz \cite{belmega-tsp-2011};
    \item for $i\in\{1,2\}$, the quantity $\mathrm{SNR}_{i}$ is, for Node $i$, the equivalent signal-to-noise ratio (SNR) at the next node after the neighbor. It is assumed to express as follows:
\tc{black}{\begin{equation}\label{snr}
\mathrm{SNR}_i=\frac{p_{i} h_{i}p'_{-i} h'_{-i}}{\sigma^{2}},
\end{equation}}
$\sigma^{2}$ being the noise variance and the \textit{index notation} $-i$ standing for the index of the other node.
\end{itemize}
\textbf{Remark 1.} The results derived in Sec.~\ref{sec:region-algorithm} hold for any utility function under the form $ u_{i}(a_0,a_{1},a_{2})$  (under some assumptions which only concern the observation structure) and not only for the specific choice made above. This choice is made to allow comparisons with existing results (and more specifically with the large set of contributions on the forwarder's dilemma) to be conducted and discussed.\\
\textbf{Remark 2.} The assumed expression of the SNR is also one possible pragmatic choice but all the analytical results derived in this paper hold for an arbitrary SNR expression of the form $\mathrm{SNR}_{i}(a_0,a_1,a_2)$; this choice is sufficiently general to study the problem of channel fluctuations which is the main feature to be accounted for. The proposed expression is relevant e.g., when nodes implement the amplify-and-forward protocol to relay the signals or packets \cite{elgamal-book-2011}. \tc{black}{This simple but reasonable model for the SNR may either be seen as an approximation where the single-hop links dominate the multi-hop links or the talk/listen phases are scheduled appropriately.} If another relaying protocol is implemented such as decode-and-forward, other expressions for the equivalent SNR may be used (see e.g., \cite{djeumou-jwcn-2006}) \tc{black}{without questioning the validity of the analytical results} provided in this paper. At last, the parameter $\alpha \geq 0$ in (\ref{eq51}) allows one to assign more or less importance to the energy consumption of the node. Indeed, as in \cite{hub}, the first term of the utility function represents the benefit of transmitting (i.e., the goodput) while the second term represents the cost of transmitting (i.e., the spent energy).

The pair of functions $(u_1, u_2)$ defines a strategic-form \textit{static game} (see e.g., \cite{lasaulce-book-2011}) in which the \textit{players} are Nodes $1$ and $2$ and the \textit{action sets} are respectively $\mc{A}_1 = \mc{P}^2$ and $\mc{A}_2= \mc{P}^2$. This game generalizes the forwarder's dilemma introduced in \cite{hub}. The latter can be retrieved by assuming that $\varphi$ is a step function, $p_i'$ is binary, $p_i$ is constant, and all the channel gains are constant. In the next section, we describe mathematically the problem under investigation. It is shown how the problem can be modeled by a repeated game, which is precisely built on the stage or static game:
\begin{equation}\label{oneshotgame}
\mc{G} = \left( \mc{N}, \{ \mc{A}_i\}_{i\in \mc{N}}, \{u_i\}_{i\in \mc{N}}   \right),
\end{equation}
where $\mc{N} = \{1,2\}$.

The unique Nash equilibrium of $\mc{G}$ is $p_{i,\mathrm{NE}} = P_{\min}$ and $p_{i',\mathrm{NE}} = P_{\min}$. If the minimum power $P_{\min}$ is taken to be zero, then the situation where the nodes do not transmit at all corresponds to the equilibrium (and thus $(u_1, u_2) = (0,0)$), which clearly shows one of the interests in modeling the packet transmission problem as a repeated game.

\section{Repeated game formulation of the problem}
 \label{sec:game-formulation}

The problem we want to solve in this paper is as follows. It is assumed that the nodes interact over \tc{black}{an infinite number of stages}. Over stage $t \in \{1,2,...,T\}$, $T \rightarrow \infty$, the channel gains are assumed to be fixed while the realizations of each channel gain are assumed to be i.i.d. from stage to stage. During a stage, a node typically exchanges many packets with its neighbors. At each stage, a node has to make a decision based on the knowledge it has. In full generality, the decision of a node consists in choosing a probability distribution over its set of possible actions. The knowledge of a node is in terms of global channel states and actions chosen by the other node. More precisely, it is assumed that Node $i\in \mc{N}$, has access to a signal which is associated with the state $a_0$ and is denoted by $s_i \in \mc{S}_i$, $|\mc{S}_i| < \infty$. At stage $t$, the observation $s_i(t) \in \mathcal{S}_i$ therefore corresponds to the \tc{black}{image (i.e., the knowledge) that Node} $i$ has about the global channel state $a_0(t)$. This signal is assumed to be the output of a memoryless observation structure \cite{elgamal-book-2011}\footnote{The memoryless assumption means that for sequences of realizations of size $t$ ($t$ being arbitrary), $\mathrm{Pr}(y_i^t | a_0^t, a_1^t, a_2^t)  = \Pi_{t'=1}^t  \mathrm{Pr} (  y_i(t') | a_0(t'), a_1(t'), a_2(t')) $.} whose conditional probability is denoted by $\daleth_i$:
\begin{equation}
 \daleth_{i}(s_{i}|a_{0}) = \mathrm{Pr}[S_i = s_i | A_0 = a_0 ],
\end{equation}
where capital letters stand for random variables whereas, small letters stand for realizations. Simple examples for $s_i$ are: $s_i = h_i$; $s_i = \widehat{h}_i$, $\widehat{h}_i$ being an estimate of $h_i$; $s_i = (h_i, h_i')$; $s_i =a_0=(h_1, h_1',h_2,h_2')$. Now, in terms of observed actions, it is assumed that Node  $i\in \mc{N}$ has imperfect monitoring. In general, Node  $i\in \mc{N}$ has access to a signal $y_i \in \mc{Y}_i$, $|\mc{Y}_i| < \infty$, which is assumed to be the output of a memoryless observation structure whose conditional probability is denoted by $\Gamma_i$:
\begin{equation}
 \Gamma_{i}(y_{i}|a_{0}, a_1, a_2) = \mathrm{Pr}[Y_i = y_i | (A_0, A_1, A_2) = (a_0, a_1, a_2)].
\end{equation}
The reason why we distinguish between the observations $s_i$ and $y_i$ comes from the assumptions made in terms of causality. Indeed, practically speaking, it is relevant to assume that a node has access to the past realizations of $s_i$ in the wide sense namely, to $s_i(1), ..., s_i(t)$ at stage $t$. However, only the past realizations in the strict sense $y_i(1), ..., y_i(t-1)$ are assumed to be known at stage $t$. Otherwise, it would mean that a node would have access to the image of its current action and that of the others before choosing the former.

At this point, it is possible to define completely the problem to be solved. The problem can be tackled by using a strategic-form game model, which is denoted by $\overline{\mc{G}}$. As for the static game $\mc{G}$ on which the repeated game model $\overline{\mc{G}}$ is built on, the \textit{set of players} is the set of nodes  $\mc{N} = \{1,2\}$. The \textit{transmission} \textit{strategy} of the Node $i$ is denoted by $\sigma_i$ and consists of a sequence of functions and is defined as follows:
\begin{equation}\label{eq:def-strategies}
    \begin{array}{lllccc}
     \sigma_{i,t}  & : & & \mathcal{S}_i^{t} \times \mc{Y}_i^{t-1}  & \to & \Delta(\mathcal{P}^2)\\
 & & & (s_i^t,  y_i^{t-1}) & \mapsto &  \pi_i(t),\\
\end{array}
\end{equation}
where:
\begin{itemize}
\item $s_i^t  = (  s_{i}(1),\ldots, s_{i}(t)  )$,     $y_i^{t-1}  = ( y_{i}(1),\ldots, y_{i}(t-1)  )$;
\item $\Delta(\mathcal{P}^2)$ represents the unit simplex namely, the set of probability distributions over the set $\mathcal{P}^2$;
\item $\pi_i(t)$ is the probability distribution used by the Node $i$ at stage $t$ to generate its action $(p_i(t), p_i'(t))$.
\end{itemize}
The type of strategies we are considering is referred to as a behavior strategy in the game theory literature, which means that at every game stage the strategy returns a probability distribution. The associated randomness not only allows one to consider strategies which are more general than pure strategies, but also to model effects such as node asynchronicity for packet transmissions. At last, the performance of a node is measured over the long run and nodes are therefore assumed to implement transmission strategies which aim at maximizing their long-term utilities.
The \textit{long-term utility function} of Node $i\in \mc{N}$ is defined as:

\begin{equation}\label{eq:utility-def}
\begin{array}{ccl}
U_i(\sigma_1,\sigma_2) &= &   \ds{\lim_{T \rightarrow \infty}}
\ds{\sum_{t=1}^{T}} \theta_t
\mathbb{E}\left[u_i(A_{0}(t),A_{1}(t),A_{2}(t))\right]\\
&=&    \ds{\lim_{T \rightarrow \infty}}
 \ds{\sum_{t=1}^{T}} \theta_t  \ds{\sum_{a_0,a_1,a_2}}  P_t(a_0,a_1,a_2)  u_i(a_{0},a_{1},a_{2}),
\end{array}
\end{equation}

where:
\begin{itemize}
\item $\sigma_i$ stands for the transmission strategy of Node $i\in \mc{N}$;
\item it is assumed that the limit in (\ref{eq:utility-def}) exists;
\item $\theta_t$ is a sequence of weights which corresponds to a convex combination that is, $0 \leq \theta_t < 1$ and $\sum_{t=1}^{\tc{black}{T}} \theta_t =1$. \tc{black}{For a repeated game with discount $\theta_t = (1-\delta) \delta^t$ and for a classical infinitely repeated game $\theta_t = \frac{1}{T}$};
\item as already mentioned, capital letters stand for random variables whereas, \tc{black}{small} letters stand for realizations. Here $A_{0}(t)$, $A_{1}(t)$, and $A_{2}(t)$ stand for the random processes corresponding to the network state and the nodes actions;
\item the notation $P_t$ stands for the joint probability distribution induced by the strategy profile $(\sigma_1,\sigma_2)$ at stage $t$.
\end{itemize}
\tc{black}{This general model thus encompasses the two well-known models for the sequence of weights which are given by the model with discount and the infinite Cesaro-mean. In the model with discount, note that} the discount factor may model different phenomena but in a wireless ad hoc network, the most relevant effect to be modeled seems to be the uncertainty that there will be a subsequent iteration of the stage game, for example, connectivity to an access point can be lost. With this interpretation in mind, the discounting factor represents e.g., the probability that the current round is not the last one \tc{black} {or, in terms of mobility, it may also represent the probability that the nodes do not move for the current stage. Therefore it may model the departure or the death of a node (e.g., due to connectivity loss) for a given routing path.} More details about these interpretation can be found in \cite{hub},\cite{lasaulce-book-2011} while \cite{neyman-ijgt-2010} provides a convincing technical analysis to sustain this probabilistic interpretation.

At this point, we have completely defined the strategic-form of the \textit{repeated game} that is, the triplet
\begin{equation}
\overline{\mc{G}} = \left( \mc{N}, \{ \Sigma_i\}_{i\in \mc{N}}, \{U_i\}_{i\in \mc{N}}   \right),
\end{equation}
where $\Sigma_i$ is the set of all possible transmission strategies for Node $i\in \mc{N}$.

One of the main objectives of this paper is to exploit the above formulation to find a globally efficient transmission scheme for the nodes. For this purpose, we will characterize long-term utility region for the problem under consideration.
It is important to mention that the characterization of the feasible utility region of a dynamic game (which includes repeated games as a special case) with an arbitrary observation structure is still an open problem  \cite{maschler}. Remarkably, as shown recently in \cite{benja}, \cite{larrousse}, the problem can be solved for an interesting class of problems. It turns out that the problem under investigation belongs to this class provided that the channel gains evolve according to the classical model of block i.i.d. realizations.

In the next section, we show how to exploit \cite{benja}, \cite{larrousse} to characterize the long-term utility region and construct a practical transmission strategy. In Sec. \ref{sec:equilibrium}, we will show how to integrate the strategic stability\footnote{\tc{black}{We will refer to the stability of a point to single deviations as strategic stability.}} property into this strategy, this property being important to ensure that selfish nodes effectively implement the efficient strategies.

\section{Long-term utility region characterization}
\label{sec:region-algorithm}

When the number of stages is assumed to be large, the random process associated with the \textit{network state} $A_0(1), A_0(2),..., A_0(T)$ is i.i.d. and, the observation structure given by $\left( \daleth_1, \daleth_2, \Gamma_1, \Gamma_2 \right)$ is memoryless, some recent results can be exploited to characterize the feasible utility region of the considered repeated game and to derive efficient transmission strategies. The main difficulty to determine the feasible utility region of $\overline{\mc{G}}$ is to find the set of possible average correlations between $a_0$, $a_1$, and $a_2$. Formally, the correlation averaged over $T$ stages is defined by:
\begin{equation}\label{average}
    P^{T}(a_0,a_1,a_2)=\frac{1}{T}\sum_{t=1}^{T} P_t(a_0,a_1,a_2),
\end{equation}
where $P_t$ is the joint probability at stage $t$. More precisely, a key notion to characterize the attainable long-term utilities is the notion of implementability, which is given as follows.
 \begin{definition}\label{def-implementability}
\textit{An average correlation $Q$ is said to be \textit{implementable} if there exists a pair of transmission strategies  $(\sigma_1, \sigma_2)$ such that the average correlation induced by these transmission strategies verifies:}
\begin{equation}
\begin{array}{l}
\forall (a_0, a_1, a_2) \in \mc{A}_0 \times \mc{A}_1 \times \mc{A}_2,\\
\displaystyle{\lim_{T\rightarrow \infty}} \frac{1}{T}\sum_{t=1}^{T} P_t(a_0,a_1,a_2) =  Q(a_0,a_1,a_2).
\end{array}
\end{equation}
\end{definition}

Using the above definition, the following key result can be proved.

\begin{Proposition}\label{proposition-utility-region} \textit{The Pareto-frontier of the achievable utility region of $\overline{\mc{G}}$ is given by all the points under the form $(\mathbb{E}_{Q_{\lambda}}(u_1), \mathbb{E}_{Q_{\lambda}}(u_2))$, $\lambda \in [0,1]$, where $Q_{\lambda}$ is a maximum point of}
\begin{equation}\label{eq:wlambda}
W_{\lambda} = \lambda \mathbb{E}_Q(u_1) + (1-\lambda) \mathbb{E}_Q(u_2),
\end{equation}
\textit{and each maximum point is taken in the set of probability distributions which factorize as follows:}
\begin{multline}\label{achie}
   Q(a_0,a_1,a_2)=\sum_{v,s_{1},s_{2}}\rho(a_{0})P_{V}(v)\times \daleth(s_{1},s_{2}|a_{0})\times \\
     P_{A_{1}|S_1, V}(a_{1}|s_1, v) P_{A_{2}|S_2, V}(a_{2}|s_2, v),
    \end{multline}
\textit{where:}
\begin{itemize}
\item  \tc{black}{\textit{$\lambda$ denotes the relative weight assigned to the utility of the first player and can be chosen arbitrarily depending on some prescribed choice e.g., in terms of fairness or global efficiency;}}
\item \textit{$\rho$ is the probability distribution of the network state $a_0$;}
\item \textit{$\daleth$ is the joint conditional probability which defines the assumed observation structure i.e., a probability which writes as}\footnote{Note that $\daleth_1$ and $\daleth_2$ are directly obtained from $\daleth$ by a simple marginalization operation.}
\begin{equation}
\daleth(s_1,s_2|a_0) = \mathrm{Pr}[(S_1,S_2) = (s_1,s_2) | A_0 = a_0 ];
\end{equation}
\item \textit{$V \in \mathcal{V}$ is an auxiliary random variable or lottery.}
\end{itemize}
\end{Proposition}
(See the proof in the appendix).
One interesting comment to be made concerns the presence of the "parameter" or auxiliary variable $V$. The presence of the auxiliary variable is quite common in information-theoretic performance analyses and in game-theoretic analyses through the notion of external correlation devices (such as those assumed to implement correlated equilibria). Indeed, $V \in \mathcal{V}$ is an auxiliary random variable or lottery which can be proved to improve the performance in general (see \cite{benja} for more details). Such a lottery may be implemented by sampling a signal which is available to all the transmitters e.g., an FM or a GPS signal.

In (\ref{achie}), $\rho$ and $\daleth$ are given. Thus, $W_{\lambda}$ has to be maximized with respect to the triplet $(P_{A_{1}|S_1, V}, P_{A_{2}|S_2, V}, P_{V})$. In this paper, we restrict our attention to the optimization of  $(P_{A_{1}|S_1, V}, P_{A_{2}|S_2})$ for a fixed lottery $P_{V}$ and leave the general case as an extension.

The maximization problem of the functional $W_{\lambda}(P_{A_{1}|S_1, V}, P_{A_{2}|S_2, V})$ with respect to $P_{A_{1}|S_1, V}$ and $P_{A_{2}|S_2, V}$ amounts to solving a bilinear program. The corresponding bilinear program can be tackled numerically by using iterative techniques such as the one proposed in \cite{Konno}, but global convergence is not guaranteed and therefore some optimality loss may be observed. Two other relevant numerical techniques have also been proposed in \cite{Gallo}. The first technique is based on a cutting plane approach while the second one consists of an enlarging polytope approach. For both techniques, convergence may also be an issue since for the first technique, no convergence result is provided and for the second technique, cycles may appear \cite{VaShe}. To guarantee convergence and manage the computational complexity issue, we propose here another numerical iterative technique namely, to exploit the sequential best-response dynamics (see e.g., \cite{fudenberg-book-1991} for a reference in the game theory literature, \cite{lasaulce-book-2011} for applications examples in the wireless area, \cite{achal} for a specific application to power control over interference channels). Here also, some efficiency loss may be observed, but it will be shown to be relatively small for the quite large set of scenarios we have considered in the numerical performance analysis. The sequential best-response dynamics applied to the considered problem translates into the following algorithm.

\textit{\textbf{Algorithm 1.}}
\begin{enumerate}
\item \textit{\textbf{Initialization}}. \textit{The arguments of the functional $W_{\lambda}$ are fixed to an initial value: $(P_{A_{1}|S_1, V}, P_{A_{2}|S_2, V})$ = $(P_{A_{1}|S_1, V}^{(0)}, P_{A_{2}|S_2, V}^{(0)})$.}

\item \textit{\textbf{Iteration}}. \textit{At iteration $n\geq 1$, $P_{A_{i}|S_i, V}^{(n)}$ is updated by being chosen in the argmax of $W_{\lambda}(P_{A_{i}|S_i, V}, P_{A_{-i}|S_{-i}, V}^{(n-1)})$. If there are several maximum points, choose one of them randomly and according to a uniform law.}

\item \textit{\textbf{Stopping criterion}}. \textit{$|  W_{\lambda}(P_{A_{i}|S_i, V}^{(n)}, P_{A_{-i}|S_{-i}, V}^{(n)}) -   W_{\lambda}(P_{A_{i}|S_i, V}^{(n-1)}, P_{A_{-i}|S_{-i}, V}^{(n-1)})   | < \eta$ for some $\eta \geq 0$.}
\end{enumerate}

Although suboptimal in general (as the available state-of-the art techniques), the proposed technique is of particular interest for at last three reasons. First, convergence is unconditional. It can be proved to be guaranteed e.g., by induction or by identifying the proposed procedure as an instance of the sequential best-response dynamics for an exact potential game (any game with a common utility is an exact potential game). Second, convergence points are local maximum points but in all the simulations performed those maximums had the virtue of not being too far from the global maximum. At last but not least, it allows us to build a practical transmission strategy which outperforms all the state-of-the art transmission strategies, as explained next. Note that this is necessarily the case when Algorithm 1 is initialized with the state-of-the art transmission strategy under consideration. \tc{black}{Although we will not tackle the classical issue of the influence of initialization on the convergence point, it is worth mentioning that many simulations have shown that the impact of the initial point on the performance at convergence is typically small, at least for the utilities under consideration. Therefore, initializing Algorithm 1 with naive strategies such as  transmitting at full power $\forall t \in \{1,...,T\}, \ (a_1(t), a_2(t)) = (P_{\max}, P_{\max}, P_{\max}, P_{\max})$ is well suited.}

\tc{black}{\textbf{Remark 3.}} Algorithm 1 would typically be implemented offline in practice. The purpose of Algorithm 1 is to generate decision functions which are exploited by the proposed transmission strategy. To implement Algorithm 1, only statistics need to be estimated in practice (namely, the channel distribution $\rho$ and the observation structure conditional distribution $\daleth$); estimating statistics such as the channel distribution information is known to be a classical issue in the communications literature.

\section{Proposed equilibrium transmission strategy}
\label{sec:equilibrium}

The main purpose of this section is to obtain globally efficient transmission strategies. Here, global efficiency is measured in terms of social welfare namely, in terms of the sum $U_1+U_2$. \tc{black}{This corresponds to choosing $\lambda=\frac{1}{2}$. This choice is pragmatic and follows to what is often done in the literature; it implicitly means that the network nodes have the same importance. Otherwise, this parameter can always be chosen to operate at the desired point of the utility region. Indeed, social welfare} is a well-known measure of efficiency and it also allows one to build other famous efficiency measure of a distributed network such as the price of anarchy \cite{papadimitriou-stoc-2001}. The proposed approach holds for any other feasible point of the utility region which is characterized in the preceding section.

The transmission strategy we propose comprises three ingredients: \tc{black}{1) a well-chosen operating point of the utility region; 2) the use of reputation
\cite{c2i7}, \cite{a2}; 3) the use of virtual credit \cite{c2i8}, \cite{c2i9}}.
\begin{enumerate}
  \item The proposed operating point is obtained by applying the sequential best-response dynamics procedure described in Sec. \ref{sec:region-algorithm} and choosing $\lambda=0.5$, $|\mc{V} |=1$. Each individual maximization operation provides a probability distribution which is denoted by $\pi_i^{\star}$. Since $W_{\lambda}$ is linear in $P_{A_{i}|S_i, V}$, the maximization operation returns a point which is one of the vertices of the unit simplex. The corresponding probability distribution has thus a particular form namely, that of a decision function under the form $f_i^{\star}(s_i)$. Therefore, when operating at this point, at stage $t$, Node $i$ chooses its action to be $a_i(t) = f_i^{\star}(s_i(t))$. This defines for Node $i$ a particular choice of a lottery over its possible actions; this lottery is denoted by $\pi_i^{\star}(t)$ and is the unit simplex of dimension $2L$ i.e., $\Delta(\mathcal{P}^2)$. By convention, the possible actions for Node $i$ are ordered according to a \textit{lexicographic ordering}. Having $\pi_i^{\star}(t) = (1, 0, 0, ..., 0) \in \Delta(\mathcal{P}^2)$ means that action $(P_1,P_1)$ is used with probability $1$ (wp1); having $\pi_i^{\star}(t)  =(0, 1, 0, ..., 0) \in \Delta(\mathcal{P}^2)$ means that action $(P_1,P_2)$ is used wp1; ...; having $\pi_i^{\star}(t)=(0, 0, 0, ..., 0, 1) \in \Delta(\mathcal{P}^2)$ means that action $(P_L,P_L)$ is used wp1.
\item The reputation (see e.g., \cite{c2i7}, \cite{c2i1}, \cite{h2}) of a neighboring node is evaluated as follows. Over each game stage duration, the nodes exchanges a certain number of packets which is denoted by $K$. This number is typically large. Since each node has access to the realizations of the signal $y_i$ for each packet, it can exploit it to evaluate the reputation of the other node at stage $t$. In this section, we assume a particular observation structure $\Gamma_1$, $\Gamma_2$, which is tailored to the considered problem of packet forwarding  in ad hoc networks. We assume that the signal $y_i$ is binary: $y_i \in \{\mathrm{D},\mathrm{F}\}$. Let $\epsilon\in [0,1]$ be a parameter which represents the \textit{probability of misdetection}. If Node $i$ chooses the action $a^{\min} = (P_{\min}, P_{\min})$ (resp. any other action of $\mc{P}_i^2$), then with probability $1-\epsilon$, Node $-i$ receives the signal $\mathrm{D}$ (resp. $\mathrm{F}$). With probability $\epsilon$, Node $-i$ perceives what we define as the action Drop $\mathrm{D}$ (resp. Forward $\mathrm{F}$) while the action Forward $\mathrm{F}$ (resp. Drop $\mathrm{D}$) has been chosen by Node $i$. \tc{black}{Thus, $\Gamma_i$ takes the following form:
    \begin{equation}\label{eq25}
    \Gamma_i(y_i|x_0,a_1,a_2)=
\left|
  \begin{array}{ll}
   1-\epsilon & \hbox{if $y_i=\mathrm{F}$ ~and~ $a_{-i} \in \mathcal{A}_i^{\mathrm{F}}$,} \\\\
   \epsilon & \hbox{otherwise,} \\\\
  \end{array}
\right.
\end{equation}
    where $\mathcal{A}_i^{\mathrm{F}}=\mc{P} _i\times \mc{P}_i \setminus \{0\}$.}

    Using these notations, Node $i$ can compute the reputation of Node $-i$ as follows:
  \begin{equation}\label{eqReputation}
    R_{-i}(t)=\frac{(1-\epsilon)|\{y_i=\mathrm{F}\}|+\epsilon |\{y_i=\mathrm{D}\}|}{K},
\end{equation}
where $|\{y_i=\mathrm{F}\}|$ and $|\{y_i=\mathrm{D}\}|$ are respectively the numbers of occurrences of the action Forward and Drop among the $K$ packets Node $i$ has been needing the assistance of Node $-i$ to forward its packets. The reputation $ R_{-i}(t)$ is one of the tools we use to implement the transmission strategy which is described further. Note that one of the interesting features of the proposed mechanism is that reputation (\ref{eqReputation}) of Node $-i$ only exploits local observations (first-hand reputation information \cite{h2}); Node $i$ does not need any information about the behavior of its neighboring nodes. This contrasts with the closest existing reputation mechanisms such as \cite{c2i7}, \cite{c2i1} for which the reputation estimation procedure exploits information obtained from other nodes (second-hand information \cite{h2}). The corresponding information exchange induces additional signaling \cite{c2i1} and additional energy consumption. At last, by using these techniques, selfish nodes may collude and disseminate false reputation values.
%
\tc{black}{\item The idea of virtual credit is assumed to be implemented with a similar approach to previous works \cite{c2i9},\cite{c2i1} namely, we assume that the nodes have an initial amount of credits, impose a cost in terms of spent credits for a node that wants to transmit through a neighbor at a certain frequency or probability, and that are rewarded when they forward their neighbors' packets. The reward and cost assumed in this paper are defined next.}
\end{enumerate}

The proposed transmission strategy is as follows. While the node has not enough credit, it adopts a cooperative decision rule, which corresponds to operating at the point we have just described. Otherwise, it adopts a signal-based tit-for-tat decision rule. Existing tit-for-tat decision rules such as GTFT \cite{c2i5} do not take into account the possible existence of a state for the game and therefore the existence of a signal associated with the realization of the state. Additionally, the proposed tit-for-tat decision rule also takes into account the fact that action monitoring is not perfect. In contrast with the conventional setup assumed to implement tit-for-tat or its variants, the action set of a node is not binary. Therefore we have to give a  meaning to tit-for-tat in the considered setup. The proposed meaning is as follows. When Node $i$ receives the signal D and Node $-i$ has effectively chosen the action Drop it means that Node $-i$ has chosen $a_{-i} = a^{\min} =  (P_{\min}, P_{\min})$.  When Node $i$ receives the signal Forward and Node $-i$ has effectively chosen the action Forward it means that Node $-i$ has chosen $a_{-i} = a_{-i}^{\star} = f_{-i}^\star(s_{-i})$. Implementing tit-for-tat for Node $i$ means choosing $a_i = a^{\min} =(P_{\min}, P_{\min})$ (which represents the counterpart of the action Drop) if the Node $-i$ is believed to have chosen the action $a_{-i} = a^{\min}=(P_{\min}, P_{\min})$. On the other hand, Node $i$ chooses the best action $a_i^{\star} = f_i^\star(s_i)$ when it perceives that Node $-i$ has chosen the action $a_{-i}^{\star} = f_{-i}^\star(s_{-i})$. Note that, contrarily to the conventional tit-for-tat decision rule, the actions $a_i^{\star}$ and $a_{-i}^{\star}$ differ in general. Denoting by $m_i(t) \geq 0$ the credit of Node $i$ at stage $t$, the proposed strategy expresses formally as follows. We will refer to this transmission strategy as SARA (for State Aware tRAnsmission strategy).\\

\textit{\textbf{Proposed transmission strategy (SARA).}}\\
\begin{equation}\label{eq25}
     \sigma_{i,t}^{\star}(s_i^t, y_i^{t-1})=
\left|
  \begin{array}{ll}
   \pi^{\star}_{i}(t) & \hbox{if $t$=0 ~or~$m_{i}(t)$$<$$\mu$,} \\\\
    \widehat{\pi}_{-i}(t-1) & \hbox{otherwise,} \\\\
  \end{array}
\right.
\end{equation}
\textit{where:}
\begin{itemize}
 %
\item \textit{the virtual credit $m_i(t)$ obeys the following evolution law:
     \begin{equation}\label{eq24}
       m_{i}(t)=  m_{i}(t-1)
       + \beta <\pi_{i}(t-1), e_k> -  \beta \nu_i(t-1),
     \end{equation}
$ \beta \nu_i(t-1) $ represents the virtual monetary cost for Node $i$ when its packet arrival rate is $\nu_i(t-1)$, with $\beta \geq 0$; $<;>$ stands for the scalar product; $e_k $ is the $k^{\mathrm{th}}$ vector of the canonical basis of $\mathbb{R}^{\tc{black}{2L}}$ namely, all components equal $0$ except the $k^{\mathrm{th}}$ component which equals $1$. \tc{black}{The index $k$ is given by the index of action $a_i^{\star}(t)=f_i^{\star}(s_i(t))$};}
\item \textit{$\mu \geq 0$ is a fixed parameter which \tc{black}{represents the cooperation level of the nodes. A sufficient condition on $\mu$ and $\beta$ to guarantee that the nodes have always enough credits is that $\mu \geq 2 \beta$;}}

\item  \textit{the distribution $\widehat{\pi}_{-i}(t-1) $ is constructed as follows:
\begin{equation} \label{estimatedact}
\widehat{\pi}_{-i}(t-1)  = R_{-i}(t-1)  \pi^{\star}_{i}(t)
+ [ 1 - R_{-i}(t-1) ] \pi^{\min},
\end{equation}
with $\pi^{\min} = (1,0,0, \ldots, 0) \in \mathbb{R}^{\tc{black}{2L}}$ representing the pure action $a^{\min}=(P_{\min}, P_{\min})$.}

\end{itemize}

\textbf{\textit{Comment 1.}} The second term of the dynamical equation which defines the credit evolution corresponds to the reward a node obtains when it forwards the packets of the other node. On the other hand, the third term is the cost paid by the node for asking the assistance of the other node to forward. The same weight (namely, $\beta$) is applied on both terms to incite the node to cooperate. Additionally, such a choice allows one \tc{black}{from preventing cooperative nodes to have an excessive number of credits, thus to avoid} using a mechanism such as in \cite{c2i1} to regulate the number of credits. \tc{black}{In \cite{c2i1}, the credit excess occurs because the reward in  terms of credits only depends on the node action and the cost only depends on the relaying node action. Thus, when the  node is cooperative and the relaying node is selfish, there will be a reward but not a cost}.
 As for the credit system, in practice it might be implemented \tc{black}{either by assuming the existence of an external central trusted entity \cite{c2i1}, \cite{tvtCred} that stores and manages the nodes credits, or} through a credit counter located in the node and maintained by a tamper-resistant security module (see e.g., \cite{c2i8}, \cite{c2i9} for more details). Thus, in practice the operation of this module would not be altered, because it would be designed so that information be accessible only by specific software containing appropriate security measures.

\tc{black}{\textbf{\textit{Comment 2.}}  Depending on whether packet forwarding rate maximization or energy minimization is sought, it is possible to tune the triplet of parameters $(m_i(0), \beta, \mu)$ according to what is desired. In this respect, it can be checked that the best packet forwarding rate is obtained by choosing any triplet under the form $(2 \beta, \beta, 2 \beta)$ for any $\beta>0$. But the power (which is defined  by the quantity ANP defined through (\ref{eq:def-ANP})) is then at its maximum. On the other hand, if the triplet of parameters takes the form $(m_i(0), 0, 0)$, with $m_i(0) \geq 0$, the consumed network power will be at its minimum and the packet forwarding rate will be minimized as well. Other choices for the triplet $(m_i(0), \beta, \mu)$ therefore lead to various tradeoffs in terms of transmission rate and consumed power.}

\textbf{\textit{Comment 3.}} The proposed strategy can always be used in practice whether or not it corresponds to an equilibrium point of $\overline{\mc{G}}$. However, if the strategic stability property is desired, some conditions have to be added to ensure that it corresponds to an equilibrium. Indeed, effectively operating at an efficient point in the presence of self-interested and autonomous nodes is possible if the latter have no interest in changing their transmission strategy. More formally, a point which possesses the property of strategic stability or Nash equilibrium is defined as follows.

 \begin{definition}\label{def-NE} \textit{A strategy profile $(\sigma_1^{\mathrm{NE}}, \sigma_2^{\mathrm{NE}})$ is a Nash equilibrium point for $\overline{\mc{G}}$ if}
 \begin{equation}
 \forall i \in \mc{N}, \forall \sigma_i \in \Sigma_i, \
 U_i(\sigma_1^{\mathrm{NE}}, \sigma_2^{\mathrm{NE}}) \geq  U_i(\sigma_i, \sigma_{-i}^{\mathrm{NE}}).
 \end{equation}
   \end{definition}

In order to obtain an explicit condition for the proposed strategy to be an equilibrium we consider, as the closest related works \cite{c2i7}, a repeated game with discount. \tc{black}{This also allows some effects such as the loss of network connectivity to be captured. Remarkably, for the repeated game model with discount, the subgame\footnote{A subgame of the repeated game is a game that starts at a stage $t$ with a given history.} perfection property is also available. This is useful in practice since it offers some robustness in terms of node behavior. Indeed, this property makes the equilibrium strategy robust against changes in terms of node behavior which might occur during the transmission process; even if some deviations from equilibrium occurred in the past, players have an interest in coming back to equilibrium. A necessary and sufficient condition for a strategy profile to be subgame perfect equilibrium is given by the following result.}

\begin{Proposition}\label{pro5} \textit{Assume that $\forall t \geq 1$, $\theta_t = (1-\delta)\delta^t$, $0 < \delta  < 1$. The strategy profile $(\sigma_{1}^{\star}, \sigma_{2}^{\star})$ defined by (\ref{eq25}) is a subgame perfect equilibrium of $\overline{\mc{G}}$ if and only if}:
 \begin{equation}\label{eq40}
\delta\geq\max\left\{\frac{c_i}{(1-2\epsilon)r_i},\frac{c_{-i}}{(1-2\epsilon)r_{-i}},0\right\},
\end{equation}
\textit{ where: $c_i=\ds{\sum_{a_0}}\rho(a_{0})(u_{i}^{1}-u_{i}^{k})$, and $r_i=\ds{\sum_{a_0}}\rho(a_{0})(u_{i}^{k}-u_{i}^{2})$. $u_{i}^{1}=u_i(a_{0},a_{1}^{\star},a^{\min})$,  $u_{i}^{k}=u_i(a_{0},a_{1}^{\star},a_{2}^{\star})$, $u_{i}^{2}=u_{i}(a_{0},a^{\min},a_2^{\star})$, and $a_i^{\star}=f_i^{\star}(s_i)$.}
\end{Proposition}
(See the proof in the appendix)

\tc{black}{\textbf{\textit{Comment 4.}} The proposed transmission strategy is compatible with a packet delivery mechanism such as an ACK/NACK mechanism. Indeed, in the definition of the transmission strategies (\ref{eq:def-strategies}), the observed signal $y_i$ may correspond to a binary feedback such as an ACK/NACK feedback. Indeed, $y_i$ corresponds to an image of $(a_0,a_1,a_2)$. Such image might then be a binary version of the receive SNR or SINR (e.g., if the receive SNR is greater than a threshold then the packet is well received and the corresponding feedback signal $y_i$ will be ACK). More generally, a binary feedback of the form Forward/Drop is completely compatible with the presence of ACK/NACK feedback-type mechanisms. Simply, the signal Drop may combine the effects of a selfish behavior and bad channel conditions.}

\textbf{\textit{Comment 5.}} This section shows that the proposed transmission strategy has five salient features.
\begin{enumerate}
\item  First of all, in contrast with the related works on the forwarder's dilemma, it is able to deal with the problem of time-varying link qualities.

\item Second, it not only deals with the adaptation of the cooperation power $p_i'$ of Node $i$ (which is the power to forward the packets of the other node) but also the power to transmit its own packets $p_i$.

\item Third, the proposed strategy is built in a way to exploit the available arbitrary knowledge about the global channel state ($a_0$) as well as possible. The key observation for this is to exploit the provided utility region characterization. Ideally, the nodes should operate on the Pareto frontier. This is possible if a suited optimal algorithm is used.

\item Fourth, the proposed transmission strategy is shown to possess the strategic stability property in games with discount under an explicit sufficient condition on the discount factor. Note that, here again, each node has only imperfect monitoring of the actions chosen by the other node. Additionally, the equilibrium strategy is subgame perfect.

\item Fifth, the proposed strategy does \tc{black}{not} induce any problem of credit outage or excess. \tc{black}{Credit outage is avoided only if the conditions $\mu\geq 2\beta$ and (\ref{eq40}) are satisfied. Therefore, if there is no credit outage problem, there is not need for assisting distant nodes as required in \cite{c2i1}}.
\end{enumerate}

\section{Numerical performance analysis}\label{sec:performance-analysis}

\tc{black}{All simulations provided in this section have been obtained by an ad hoc simulator developed under \textit{Matlab}.} The simulation setup we consider in this paper is very close to those assumed in the closest works and \cite{c2i7}, \cite{c2i9}, \cite{c2i1} in particular. The setup we assume by default is provided in Sec. \ref{sec:setup}. When other values for some parameters are considered, this will be explicitly mentioned. In addition to the simulation setup subsection, the simulation section comprises three subsections. The first subsection (Sec. \ref{sec:utilana}) aims at conducting a performance analysis in terms of utility function (\ref{eq51}), which captures the tradeoff between the transmission rate and the energy spent for transmitting.  Sec. \ref{sec:packetana} focuses on the transmission rate aspect  while Sec. \ref{sec:energyana} is dedicated to a performance analysis in terms of consumed network energy.

\subsection{Simulation setup assumed by default}\label{sec:setup}

We consider a network of $N$ nodes. When $N$ is considered to be fixed, it will be taken to be equal to $50$. The $N$ nodes are randomly placed (according to a uniform probability distribution) over an area of $1000 \times 1000$ $\mathrm{m}^2$; \tc{black}{only network topology draws which guarantee every node to have a neighbor (in the sense of its radio range) are kept}. \tc{black}{The assumed topology corresponds to a random topology since the node locations are drawn from a given spatial distribution law (which is uniform for the simulations). Each node only considers the behavior of its neighbors to choose its own behavior. As assumed in the related literature, if a node has several neighbors, it is assumed to play a given game with each of its neighbors. In fact, averaging the results over the network topology realizations has the advantage of making the conclusions less topology-dependent.} Provided simulations are averaged over $1200$ draws for the network topology. \tc{black}{Routes are supposed to be fixed and known.} \tc{black}{Indeed, the proposed transmission strategy is compatible with any routing algorithm.} One node can communicate with another node only if the inter-node distance is less than the radio range, which is taken to be $150$ m. When a node has several neighbors, \tc{black}{it may be involved in several routing paths}, then it is assumed to play several \tc{black}{independent} forwarding games in parallel, \tc{black}{and have a given initial credit $m_i(0)$ for each neighbor. The credits are updated separately based on the corresponding forwarding game. This means that the credits a node receives by cooperating with one of its neighbors can only be used for forwarding via the considered neighbor. As a node without neighbors does not need credits and the nodes do not obtain an initial credit, the problem of  credit excess is avoided}. By default, $50\%$ of the nodes are assumed to be selfish but the network does not comprise any malicious node. The initial packet forwarding rate for cooperative nodes and selfish nodes are respectively set to $1$ and $0.1$. Each source node transmits at a constant bit rate of $2$ packets/s. For each draw for the network topology, the simulation is run for $1000$ s. This period of time is made of $20$ frames of $50$ s. A frame corresponds to a game stage and to a given draw for the channel vector $h$. \tc{black}{The fact that channel gains are assumed to fluctuate over time is a way of accounting for mobility; in the simulations, they are assumed to evolve according to a (discrete version of the) Rayleigh fading law. Averaging over network topologies allows one to average the results over the path losses.} Each channel gain is thus drawn according to an exponential law, which corresponds to a Rayleigh law for the amplitude; if one denotes by $h_i$ the considered channel gain, we have that $h_i \sim \frac{1}{\overline{h}_i}  e^{-\frac{h_i}{\overline{h}_i}}$, where $\overline{h}_i = \mathbb{E}(h_i)$ represents the path loss effects. \tc{black}{As mentioned above, the channel gain is discrete and the discrete realizations are obtained by quantizing the realizations given by a Rayleigh distribution. The effect of quantization on the performance is typically small. Simulations, which are provided here, show that the loss induced by implementing Algorithm 1 by using quantized channel gains in the presence of actual channel gains which are continuous is about a few percents for the size of channel gain sets used for the simulations.} If $d$ denotes the inter-node distance of the considered pair of nodes, then the path loss is assumed to depend on the distance according to $\overline{h}_i = \frac{\mathrm{const}}{d^2+\kappa^2}$; $\kappa>0$ is a distance which is used to avoid numerical divergence in $d=0$. In practice, $\kappa$ may typically represent the antenna height. During each frame, $100$ packets are exchanged. Tab. \ref{tab1} recaps the values chosen for the main network parameters.

\begin{table}[htbp]
\begin{center}
\caption{Simulation Settings}  \label{tab1}
\begin{tabular}{l|c}
\hline
Network parameters & Value\\
  \hline
  Space & $1000$ m $\times$ $1000$ m\\
  Number of nodes & $N=50$ \\
  Radio range & $150$ m \\
  Const & $10^3$ \\
  $\kappa$ & $5$ m \\
  Initial credit  & $m_i(0)=35$ \\
  \tc{black}{Initial credit of ICARUS}  \cite{c2i1}& $m_i(0)=220$ \\
  Parameter cost & $\beta = 10$ \\
  Cooperation degree & $\mu = 20$ \\
  \tc{black}{Probability of misdetection} & \tc{black}{$\epsilon =0$} \\
  \tc{black}{Packet arrival rate} & \tc{black}{$\nu=1$} \\
  Simulation time  & $1000$ s\\
  Frame/stage duration & $50$ s\\
  Generosity parameter of GTFT & $0.1$ \\
  IFN of ICARUS \cite{c2i1} & $5$ \\
   $\mathrm{edp}_{\mathrm{th}}$ of ICARUS \cite{c2i1} & $0.85$ \\
   $a$ of ICARUS \cite{c2i1} & $0.5$ \\
   $b$ of ICARUS \cite{c2i1} & $2.3$ \\
  Number of topology draws & $1200$\\
  \hline
\end{tabular}
\end{center}
\end{table}

Concerning the game parameters, the following choices are made by default. The parameter $\alpha$ is set to $10^{-2} $. The receive variance $\sigma^2$ is always set to $0.1$. The sets of possible power levels are defined by: $\forall i \in \{1,2\}, \mathcal{P}_i =\mathcal{P}_i' $, $L=10$, $P_{\min}=0$, $P_{\max}= 10$ W. The power increment is uniform over a dB scale, starting from the minimal positive power which is taken to be equal to $10$ mW. The sets of possible channel gains are defined by: $\forall i \in \{1,2\}, \mathcal{H}_i =\mathcal{H}_i' $: $H=10$,
$h_{\min} = 0.04$, $h_{\max} = 10$, and the channel gain increment equals $\frac{10-0.04}{10}$. The different means of the channel gains are given by: $(\bar{h}_i,\bar{h'}_i,\bar{h}_{-i},\bar{h'}_{-i})=(1,1,1,1)$. The communication efficiency function is chosen as in \cite{belmega-tsp-2011}:  $\varphi(x) = e^{-\frac{c}{x}}$ with $c=2^r -1$, $r$ being the spectral efficiency in bit/s per Hz  \cite{belmega-tsp-2011}. In the simulations provided we always have $r=1$ bit/s per Hz; one simulation will assume a higher spectral efficiency namely, $r=3$ bit/s per Hz.

\subsection{Utility analysis}\label{sec:utilana}

Here, to be able to easily represent the utility region for the considered problem and to be able to compare our approach with previous models (with the well-known forwarder's dilemma model \cite{hub} in particular), we consider two neighboring nodes. 

\begin{figure}[htbp]
\begin{center}
  \includegraphics[width=8.5cm,height=6cm]{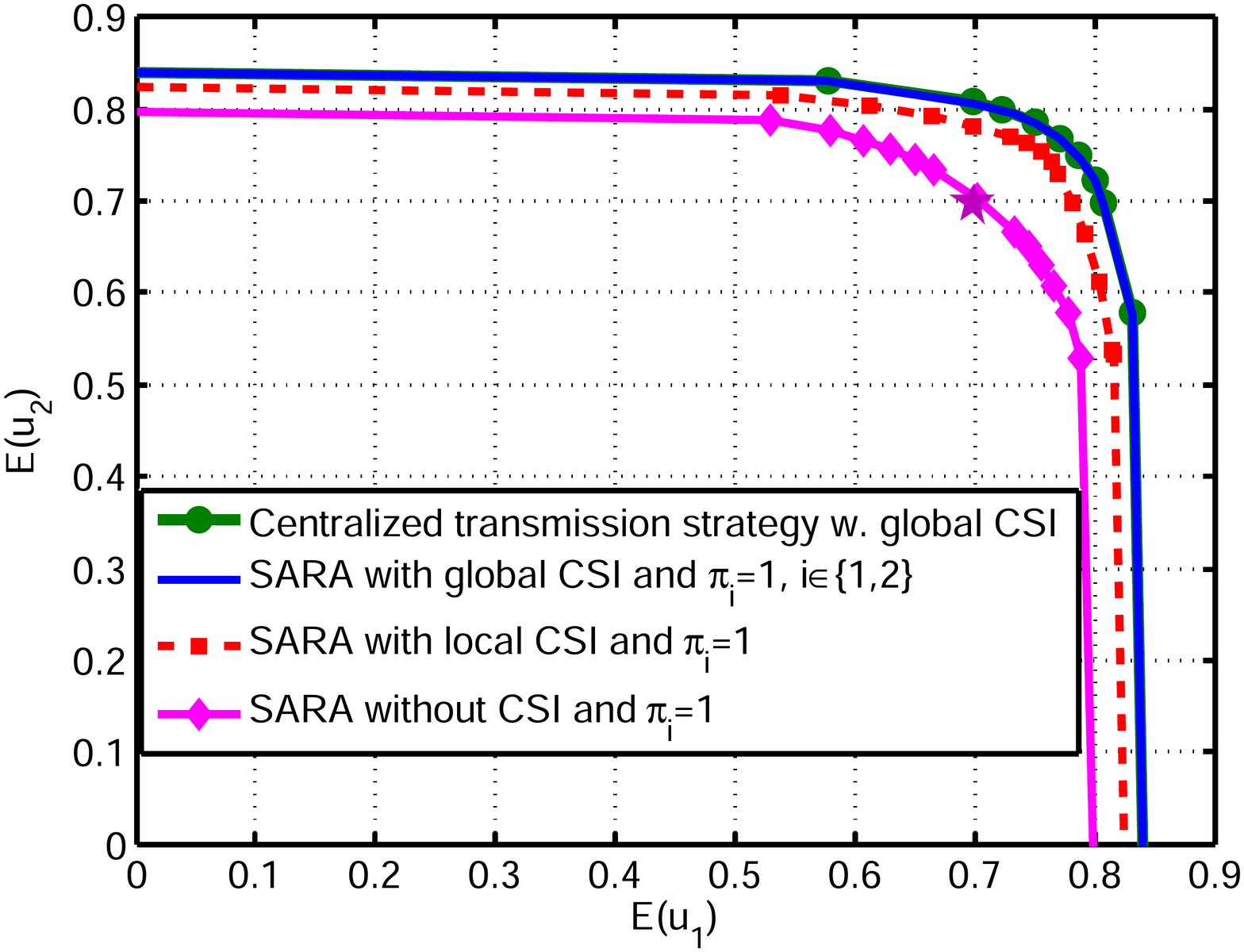}
  \end{center}
\caption{\tc{black}{Achievable utility region under various scenarios of information for the nodes: global CSI, local CSI, and no CSI.}}
\label{fig1b}
\end{figure}

The first question we want to answer is to know to what extent the ability for a node to properly adapt to the link qualities which have an impact on the weighted utility $w_{\lambda} = \lambda u_1 + (1-\lambda) u_2$, it is related to its knowledge about these qualities i.e., the global channel state $a_0=(h_1,h_1',h_2,h_2')$. To this end, we have represented in \textbf{Fig. \ref{fig1b}}, the achievable utility region under various information assumptions. The top curve in solid line represents the Pareto frontier which is obtained when implementing the transmission strategy given by Algorithm 1 when $\forall i \in \mc{N}$, $s_i=a_0=(h_1,h_1',h_2,h_2')$ that is, each node has \textit{global CSI}. The crosses correspond to the performance of the \textit{centralized transmission strategy}, namely the best performance possible. It is seen that for a typical scenario the proposed algorithm does not involve any optimality loss. The curve in dashed line is obtained with Algorithm 1 under \textit{local CSI} i.e., $s_i = (h_i, h_i')$. Interestingly, the loss for moving from global CSI to local CSI is relatively small. This shows that it is possible to implement a distributed transmission strategy without sacrificing too much the global performance. This result is not obvious since the weighted utility $w_{\lambda}$ depends on the whole vector $a_0$. When \textit{no CSI} is available (i.e., $s_i = \mathrm{constant}$), the incurred loss is more significant. Indeed, the curve with diamonds (which is obtained by choosing for each $\lambda \in [0,1]$ the best action profile in terms of the expected weighted utility\footnote{We therefore assume that the corresponding statistical knowledge is available and exploited.} (\ref{eq:wlambda})) shows that the gain in terms of sum-utility or social welfare when moving from no CSI to global CSI is about $10\%$. The point marked by a star indicates the operating point for which transmitting at full power $a_i = (P_{\max}, P_{\max}, P_{\max}, P_{\max})$ is optimal under no CSI.

\begin{figure}[htbp]
\begin{center}
  \includegraphics[width=8.5cm,height=6cm]{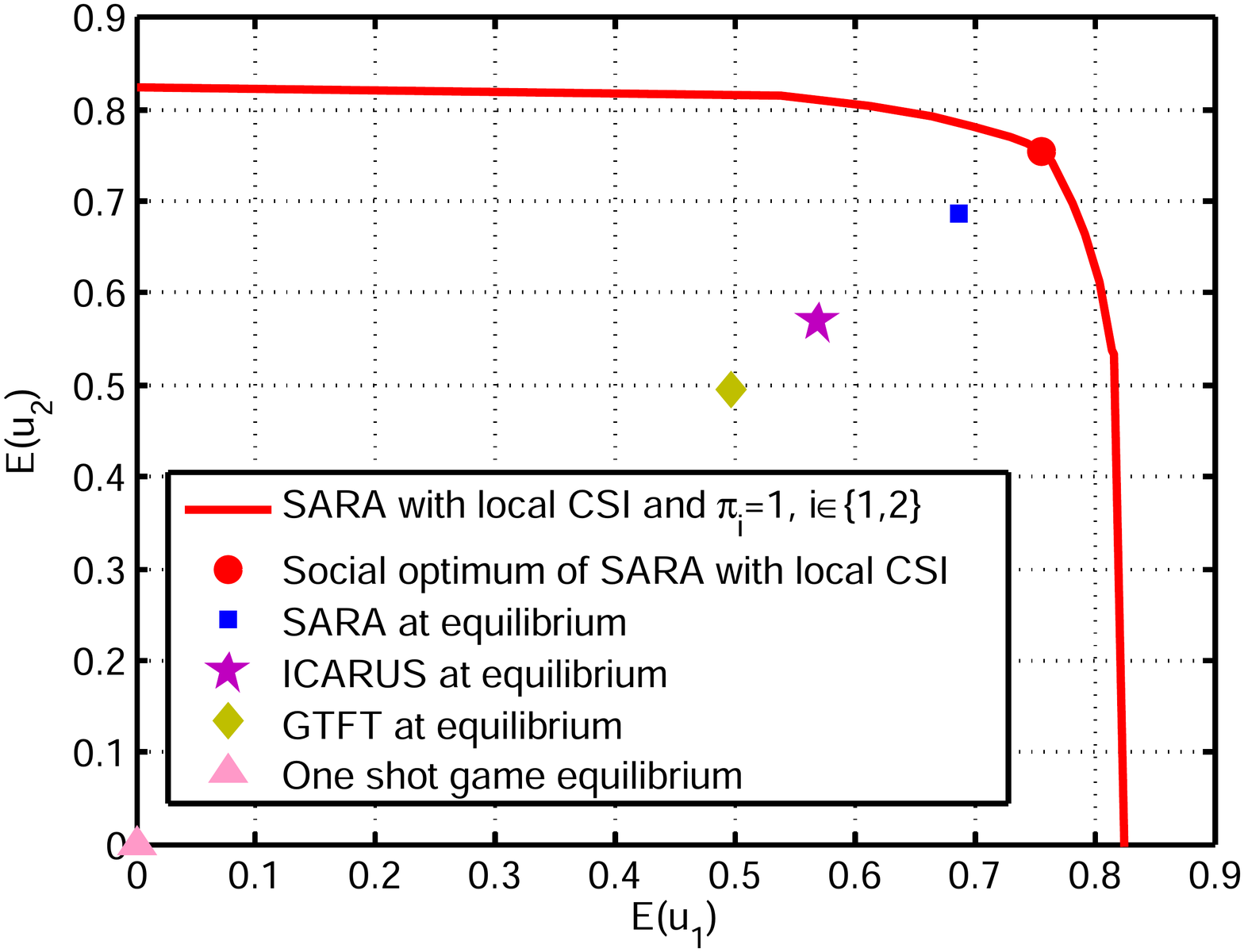}
  \end{center}
\caption{Achievable utility region with local CSI and the repeated game equilibrium for each strategy: SARA, ICARUS, and GTFT. The one-shot game Nash equilibrium is also represented. \tc{black}{The strategies ICARUS, and GTFT do not take into account the channel fluctuations.}}
\label{fig2}
\end{figure}

As a second step, we compare the performance of SARA, ICARUS \cite{c2i1} and GTFT \cite{c2i5}, that do not take into account the channel fluctuations. The three corresponding equilibrium points are particular points of the achievable or feasible utility region represented by \textbf{Fig. \ref{fig2}}. The outer curve is the achieved utility region of Fig. \ref{fig1b} when Algorithm 1 is implemented under local CSI (it is the same as the dashed line curve of Fig. \ref{fig1b}). \tc{black}{The social optimum corresponds to the point indicated by the small disk. The point marked by a square corresponds to the performance of SARA whereas, the points marked by a star and a diamond respectively represent the equilibrium points obtained when using ICARUS \cite{c2i1} and GTFT \cite{c2i5}}. Note that the way the strategies ICARUS and GTFT have been designed is such that they are able to adapt the packet forwarding rates but not the transmit power level. As a consequence they cannot exploit any available knowledge in terms of CSI, which induces a quite significant performance loss; it is assumed that GTFT and ICARUS use a pair of actions $(a_1,a_2)$ which maximizes the expected sum-utility. The gain obtained by the proposed transmission strategy comes not only from the fact that the transmit power level can adapt to the wireless link quality fluctuations, but also from the proposed cooperation mechanism. The latter both exploits the idea of virtual credit and reputation, which allows one to obtain a better packet forwarding rate than ICARUS and GTFT. We elaborate more on this aspect in the next subsection. At last, when implementing a transmission strategy built from the one-shot game model given in Sec. \ref{sec:System-model}, the NE of \cite{hub} would be obtained i.e., the operating point would be $(0,0)$, which is very inefficient.
\vspace{-0.4cm}

\subsection{Packet forwarding rate analysis}\label{sec:packetana}
In the previous subsection we have been assessing the benefits from implementing the proposed transmission strategy in terms of utility. The utility implements a tradeoff between the transmission rate and the consumed energy. Here, we want to know how good is the proposed strategy in terms of packet forwarding rate that is, the packet transmission probability.

\begin{figure}[h!]
\begin{center}
\includegraphics[width=8.5cm,height=6cm]{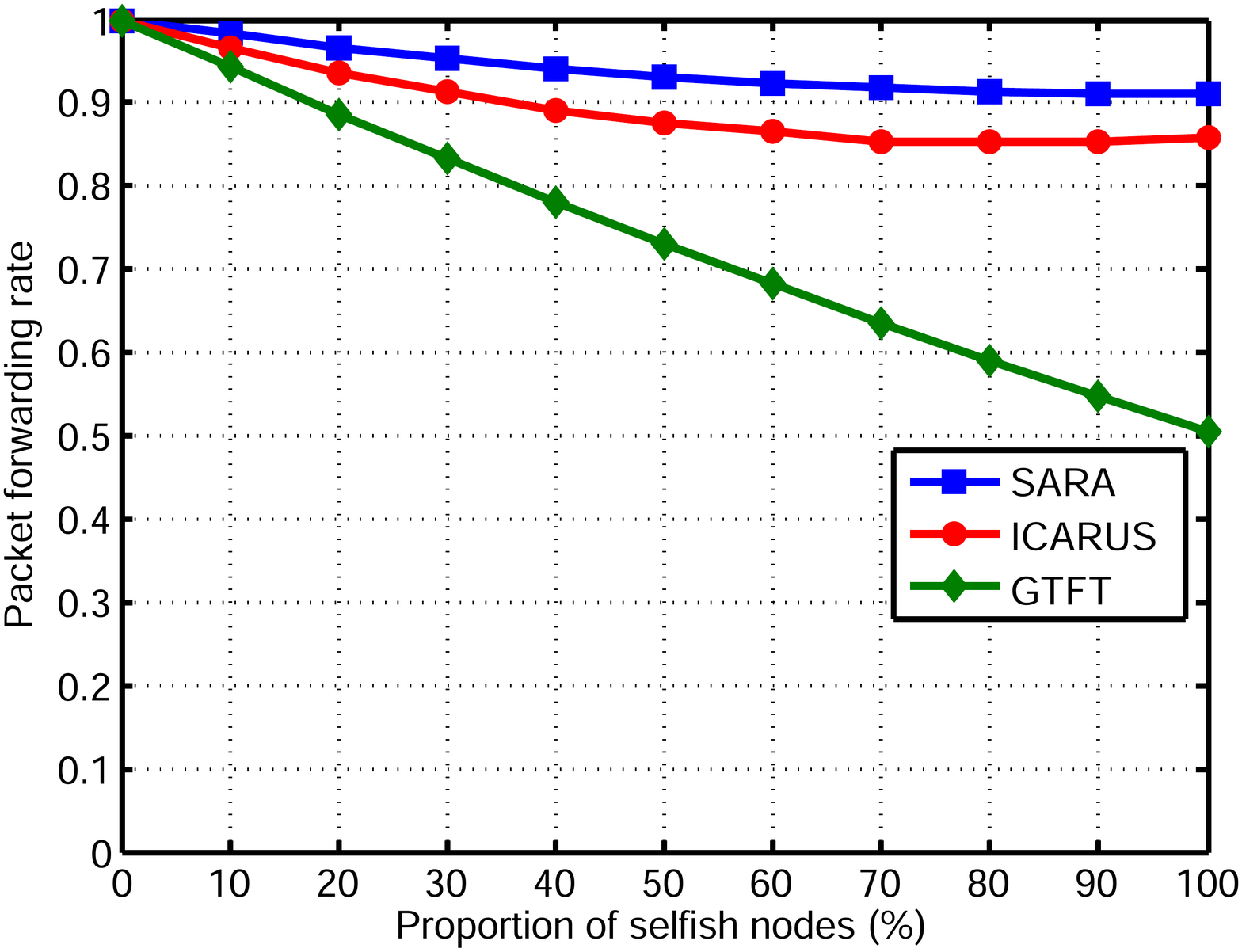}\\
\end{center}
\caption{Packet forwarding rate for different values for the proportion of selfish nodes. The results are averaged over $1200$ executions, using the simulation setup assumed by default.}\label{fig8}
\end{figure}

\textbf{Fig. \ref{fig8}} depicts the evolution of the packet forwarding rate for SARA, ICARUS, and GTFT \tc{black}{for a network of $50$ nodes}. We look at the influence of the fraction of selfish nodes. SARA is very robust to selfishness. Whatever the fraction of selfish nodes, SARA provides a high performance in terms of packet forwarding rate. We see that ICARUS is less efficient than SARA in terms of stimulating cooperation in the presence of selfish nodes, which shows that the proposed punishment mechanism is effectively relevant. The GTFT strategy performance decreases in a significant manner with the number of selfish nodes. For the latter transmission strategy, it is seen that, when the network is purely selfish, the operating packet forwarding rate is about $50 \%$; this shows the significant loss induced by using a cooperation scheme which is not very robust to selfishness.

\begin{figure}[h!]
\begin{center}
\includegraphics[width=7.5cm,height=6cm]{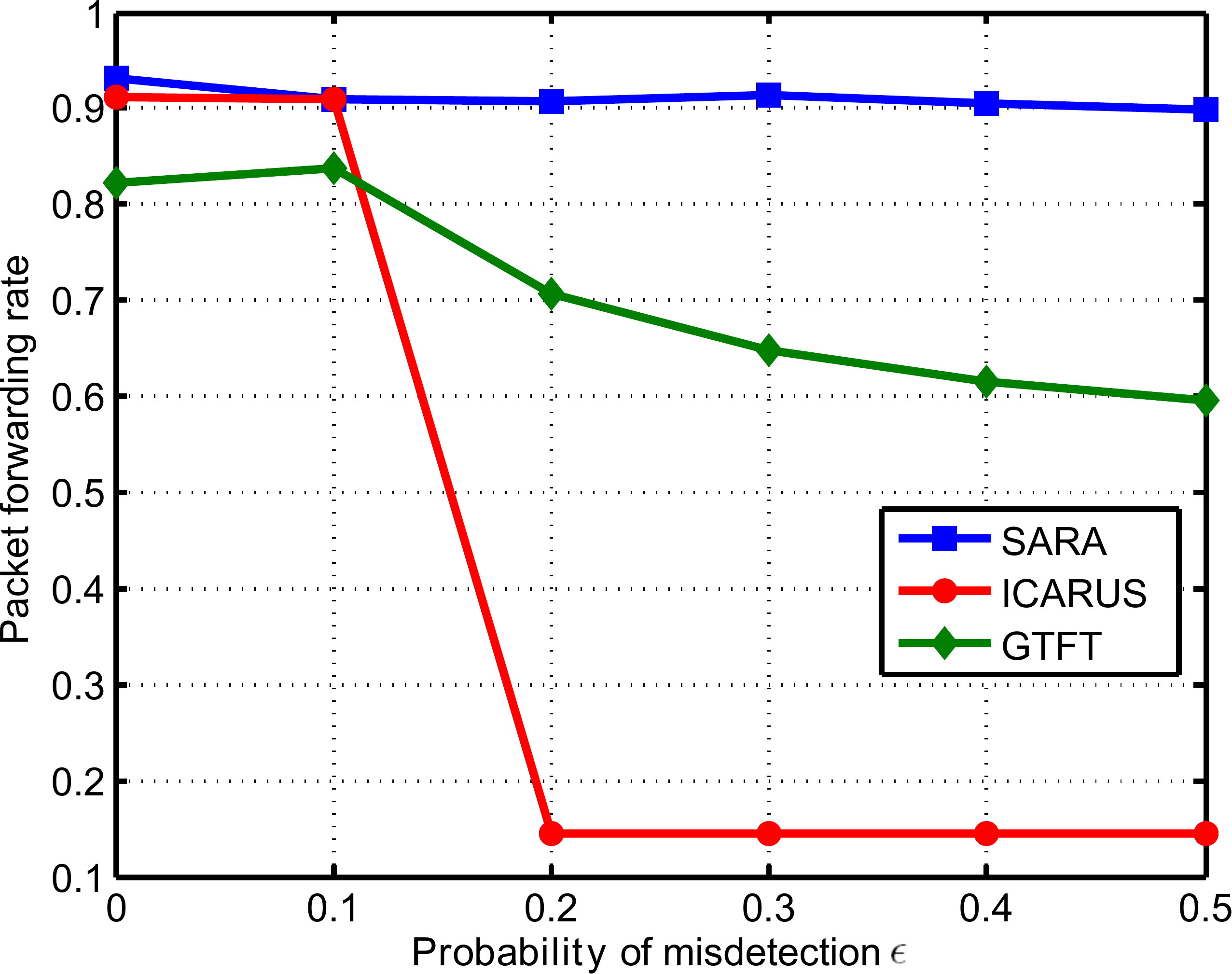}\\
\end{center}
\caption{Packet forwarding rate for different values for the probability of misdetection $\epsilon$. The results are averaged over $1200$ executions, using the simulation setup assumed by default.}\label{fig9}
\end{figure}

 The robustness to observation errors is assessed. More precisely, we want to evaluate the impact of not observing the action Forward or Drop perfectly on the packet forwarding rate. \textbf{Fig. \ref{fig9}} depicts the packet forwarding rate as a function of the probability of misdetection $\epsilon$ (see (\ref{eqReputation})). When $\epsilon > 10 \% $, the performance of ICARUS sharply decreases. This is because the retaliation aspect becomes a dominant effect. Nodes punish each other whereas, they should not; this is due to the fact that the estimates of the forwarding probabilities become poor and the ICARUS mechanism is sensitive to estimation errors; illegitimate punishments are implemented, leading to a very inefficient network. On the other hand, observation errors have little influence on SARA because under the equilibrium condition, provided that the credit is less than $\mu$, nodes keep on cooperating. Estimating the forwarding rate does not intervene in the decision process of a node. Note that we have only considered $\epsilon \leq 50\% $. The reason for this is as follows. When $\epsilon > 50 \% $ it is always possible, by symmetry, to decrease the effective probability of misdetection to $\epsilon' = 1 - \epsilon$. For this, it suffices to declare the used action to be Forward whereas, the action Drop was observed, and vice-versa.


\subsection{Consumed network energy analysis}\label{sec:energyana}
\begin{figure}[h!]
\begin{center}
\includegraphics[width=8.5cm,height=6cm]{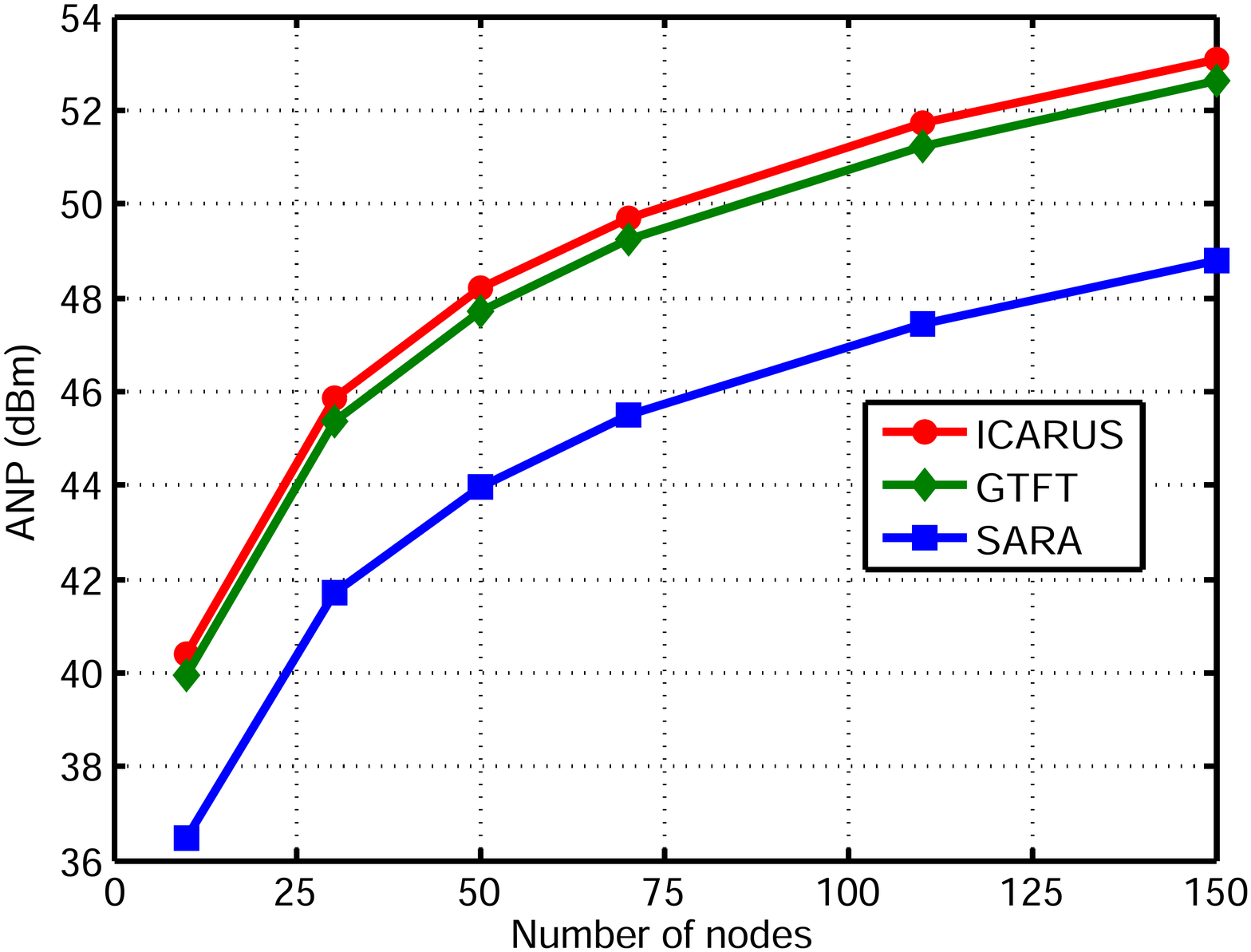}\\
\end{center}
\caption{The figure depicts the average total power (in dBm) or equivalently the total energy consumed by the network against the total number of nodes in the network. It is seen that SARA allows the energy consumed by the network to be divided by more than $2$ when compared to the state-of-the art strategies \tc{black}{ ICARUS, and GTFT that do not take into account the channel fluctuations.}}\label{fig10}
\end{figure}

Based on the preceding two sections, we know that SARA provides improvements in terms of utility and packet forwarding rate. But the most significant improvements are in fact obtained in terms of consumed energy, \tc{black} {which is also observed  in \cite{tvtPC}, where time-varying channels are exploited to find the optimal instants to communicate}. Indeed, ICARUS and GTFT have not been designed to account for link quality fluctuations whereas, SARA both adapts the packet forwarding rate and the transmit power level \tc{black}{using the parameters assumed by default, except for the path loss $\overline{h}_i = \frac{\mathrm{const}}{d^2+\kappa^2}$, where const=$10^3$ and $\kappa=5$}. In the current formulation of ICARUS and GTFT, the transmit power is fixed (as in Sec. \ref{sec:utilana}) according to the best pair of actions in terms of expected sum-utility. In this subsection, the advantage of adapting the power to the quality of the wireless link is clearly observed. Since the consumed network energy is proportional to the network sum-power averaged over time, we will work with the \textit{average network power} (ANP). Here we consider the total power which is effectively consumed by the node and not the radiated powers $p_i$ and $p_i'$ (the consumed power therefore includes the circuit power in particular). As explained e.g., in \cite{betz-tsp-2008}, \cite{richter-vtc-2009}, \cite{varma-tvt-2015}, a reasonable and simple model for relating the radiated power and the consumed power is the affine model: $p_{i,\text{total}} = a (p_i+p_i') + b$. The parameter $b$ is very important since it corresponds to the power consumed by the node when no packet is transmitted; in \cite{richter-vtc-2009}, \cite{varma-tvt-2015} it represents the circuit power whereas, in \cite{betz-tsp-2008} it represents the node computation power. Here we assume as in \cite{varma-tvt-2015} that $b$ is comparable to the $P_{\max}$ and choose the same typical values as in \cite{varma-tvt-2015} namely, $b=P_{\max}= 1 $ W. Eventually, the ANP is obtained by averaging the following quantity $\sum_{i=1}^N \left\{ a  [p_i(t) + p_i'(t)] \pi_i(t)  + b \right\}$ over all channel and network topology realizations, where $N$ is always the number of nodes in the network and $\pi_i(t)$ the forwarding probability for stage or frame $t$:
\begin{equation}\label{eq:def-ANP}
\mathrm{ANP} = \frac{1}{T'} \sum_{i=1}^N \left\{ a  [p_i(t) + p_i'(t)] \pi_i(t) + b\right\}
\end{equation}
where $T'$ corresponds to the number of realizations used for averaging. Here, this quantity is averaged over $1200 \times 20$ stages, the number of network realizations being $1200$ and the number of channel realizations being $20$. \textbf{Fig. \ref{fig10}} shows how the ANP in dBm scales with the number of nodes for SARA, ICARUS, and GTFT. It is seen that the ANP and therefore the total energy consumed by the network can be divided by more than $2$ (the gain is about $4$ dB to be more precise) showing the importance of addressing the problem of packet forwarding and power control jointly.

\subsection{Impact of quantizing channel gains}

As motivated in the System Model section, one strong argument for assuming discrete channel gains is that, technically, it corresponds to the most general case; the continuous case follows by using standard information-theoretic arguments. But, from a practical point of view, it matters to assess the loss induced by using an algorithm which exploits quantized channel gains instead of continuous ones. Fig.~XXX represents the performance in terms of XXX as a function of XXX. It is seen that the performance obtained by using discrete channel gains in the proposed algorithm whereas, the actual channel gains are continuous is typically small. Here the parameters have been chosen as follows: XXX....

\section{Conclusion}\label{sec7}

One of the contributions of this work is to generalize the famous and insightful model of the forwarder's dilemma \cite{hub} by accounting for channel gain fluctuations. \tc{black}{Therefore, the problem of knowledge about global channel state appears, in addition to the problem of imperfect action monitoring when the interaction is repeated. In this paper,} we have seen that it is possible to characterize the best performance of the studied system even in the presence of partial information; the corresponding observation structure is arbitrary provided. The observations are generated by discrete observation structures denoted by $\daleth$ and $\Gamma$. In terms of performance, designing power control policies which exploit \tc{black}{the available knowledge as well as possible} is shown to lead to significant gains. Since, we are in presence of selfish nodes, we propose a \tc{black}{mechanism to stimulate cooperation among nodes}. The proposed mechanism is both reputation-based and credit-based. For the reputation aspect, one of the novel features of the proposed strategy is that it generalizes the concept of tit-for-tat to a context where actions are not necessarily binary. For the credit aspect, we propose an evolution law for the credit which is shown to be efficient and robust to selfishness and especially imperfect action monitoring.

From the quantitative aspect, the proposed transmission strategy (referred to as SARA) Pareto-dominates ICARUS and GTFT for the utility, the packet forwarding rate, and the energy consumed by the network. Significant gains have been observed; one very convincing result is that the energy consumed by the network can be divided by $2$ when the packet forwarding problem and the power control problem are addressed jointly. 

This paper provides the characterization of the best performance in terms of transmission strategy under partial information. Although all performed simulations show that the optimality loss appears to be small, there is no guarantee that the proposed algorithm provides an optimal solution of the optimization problem to be solved to operate on the Pareto-frontier or the utility region. Providing such a guarantee would constitute a valuable extension of the present work. Another significant extension would
be to relax the i.i.d. assumption on the network state. In this
work, the network state corresponds to the global channel state
and the i.i.d. assumption is known to be reasonable but,
in other setups, where the state represents e.g., a queue length, a buffer size, or a battery level, the used framework would need to be extended since Markov decision processes would be involved.

\section*{Appendix A: Proof of Proposition 1}

First, it has to be noticed that long-term utilities are linear images of the implementable distribution. Therefore, characterizing the achievable utility region amounts to characterizing the set of implementable distributions. \tc{black}{Note that the set of implementable characterization does not depend on the assumed choice for the infinite sequence of weights $(\theta_t)_{t \geq 1}$, making the result valid for both considered models of repeated games (namely, the classical infinitely repeated game and the model with discount).}

Second, to obtain the set of implementable distributions we exploit the implementability theorem derived in \cite{benja}. Therein, it is proved that under the main assumptions of the present paper (namely, the network state is i.i.d. and the observation structure is memoryless) a joint distribution is implementable if and only if it factorizes as in (\ref{achie}). That is, a joint probability distribution or correlation $Q(a_0, a_1, a_2)$ is implementable if and only if it factorizes as:
\begin{multline}\label{achie}
   Q(a_0,a_1,a_2)=\sum_{v,s_{1},s_{2}}\rho(a_{0})P_{V}(v)\times \daleth(s_{1},s_{2}|a_{0})\times \\
     P_{A_{1}|S_1, V}(a_{1}|s_1, v) P_{A_{2}|S_2, V}(a_{2}|s_2, v).
    \end{multline}

Third, a key observation to be made now is that if two probability distributions $Q_1$ and $Q_2$ are implementable, then the convex combination $\mu Q_1 + (1-\mu) Q_2$ is implementable. Indeed, if there is a transmission strategy to implement $Q_1$ and an other to implement $Q_2$ then by using the first one $\frac{T_1}{T}$ of the time and the second one  $\frac{T-T_1}{T}$ of the time, and making $T_1$ large such that  $\frac{T_1}{T} \rightarrow \mu$, $\mu Q_1 + (1-\mu) Q_2$ becomes implementable. It follows that the long-term utility region is convex. Therefore, the Pareto frontier of the utility region, which characterizes the utility region, can be obtained by maximizing the weighted utility $W_{\lambda}$. This concludes the proof.

\section*{Appendix B: Proof of Proposition 2}

\begin{proof}
We want to prove the following result.\\

The strategy profile $(\sigma^{\star}_{i},\sigma^{\star}_{-i})$ is a subgame perfect equilibrium of $\bar{\mathcal{G}}$ if and only if:
 \begin{equation}\label{append1}
\delta\geq\max\left\{\frac{c_i}{(1-2\epsilon)r_i},\frac{c_{-i}}{(1-2\epsilon)r_{-i}},0\right\},
\end{equation}
 where: $c_i=\ds{\sum_{a_0}}\rho(a_{0})(u_{i}^{1}-u_{i}^{k})$, and $r_i=\ds{\sum_{a_0}}\rho(a_{0})(u_{i}^{k}-u_{i}^{2})$.\\

$u_{i}^{1}=u_i(a_{0},a_{1}^{\star},a^{\min})$, $u_{i}^{k}=u_i(a_{0},a_{1}^{\star},a_{2}^{\star})$, $u_{i}^{2}=u_{i}(a_{0},a^{\min},a_2^{\star})$, and $a_i^{\star}=f_i^{\star}(s_i)$.

As a preliminary, we first review the one-shot deviation "principle" in the context of interest. This "principle" is one of the elements used to prove the desired result.

\textbf{One-shot deviation principle:} For Node $i$ the one-shot deviation principle from strategy $\sigma_i$ is a strategy $\tilde{\sigma}_i$ writes as:
   \begin{equation}\label{eq35}
     \exists!~ \tau,~\forall~~t\neq \tau ~~~\sigma_{i,t}(s_i^t,y_i^{t-1})=\tilde{\sigma}_{i,t}(s_i^t,y_i^{t-1}).
   \end{equation}

The two strategies $\tilde{\sigma_{i}}$ and $\sigma_{i}$ therefore produce identical actions except at stage $\tau$. 
   \begin{definition}\label{def4}
For Node $i$ the one-shot deviation $\tilde{\sigma}_i$ from strategy $\sigma_{i}$ is not profitable if:
 \begin{equation}\label{eq36}
    U_{i}(\sigma_i,\sigma_{-i}) \geq U_{i}(\tilde{\sigma}_{i}, \sigma_{-i}),
 \end{equation}
with $\tilde{\sigma}_{i}\neq \sigma_{i}.$
   \end{definition}

 Let us exploit the one-shot deviation principle to prove the result, \tc{black}{since it is well known that a strategy profile $\sigma$ is a subgame perfect equilibrium if and only if there are no profitable one-shot deviations}. Assume that for a given game history, the distributions used by Nodes $i$  and $-i$ are respectively $\pi_i$ and $\pi_{-i}$. Following the proposed strategy $\sigma^{\star}_{i,t}$ defined by (15), and by using (14) and (17), one can obtain the distribution of a node $i$, $\pi_i(t)$, from $\pi_{-i}$  for each stage $t$. As defined by relation (17), at each stage $t$, if $m_i(t)\geq\mu$, Node $i$ chooses a distribution $\pi_{i}(t)=\hat{\pi}_{-i}(t-1)$ stipulating that $a^{min}=(P_{\min},P_{\min})$ and $a^{\star}_i(t)=f_i^{\star}(s_i(t))$ are the only actions that could be chosen with a positive probability. Thus, it would be sufficient to provide only the $k^{th}$ component of $\pi_i(t)$, which is denoted by $\pi_i^k(t)$. Note that $k$ is the index of action $a_i^{\star}(t)=f_i^{\star}(s_i(t))$. 

Thus, for $t\geq 1$ we have that:
\vspace{-0.2cm}
$$
    \pi^k_{i}(t)=
\left\{
  \begin{array}{ll}
    1, & \hbox{if $m_{i}(t)$$<$$\mu$;}\\\\
     (1-2\epsilon)^t\pi_{-i}+\epsilon\sum_{k=0}^{t-1}(1-2\epsilon)^k, & \hbox{if $mod(t,2)=1;$}\\\\
     (1-2\epsilon)^t\pi_{i}+\epsilon\sum_{k=0}^{t-1}(1-2\epsilon)^k, & \hbox{if $mod(t,2)=0.$}
  \end{array}
\right.
$$
Now, we define a one-shot deviation. We consider that Node $i$ deviates unilaterally at one stage from the proposed strategy $\sigma^{\star}_{i,t}$, by choosing $\tilde{\sigma}_{i,t}$. If Node $i$ deviates, it will be in order to save energy, thus it chooses $a^{min}$ with a higher probability than the one provided by the proposed strategy $\sigma^{\star}_{i,t}$. Therefore, we consider that $\tilde{\sigma}_{i,t}$ defines a distribution over the action set as follows $(\ref{devstra})$:
\begin{equation}\label{devstra}
    \tilde{\pi}_i(t) = \pi_i(t)-d.(\underbrace{-1}_{a^{min}},0,\ldots,\underbrace{1}_{a^{\star}},0,\ldots),
\end{equation}

where: $d\in[0,1]$, and $\pi_i(t)$ is the distribution given by $\sigma^{\star}_{i,t}$, and whose $k^{th}$ component is defined above.

Using the one-shot deviation $\tilde{\pi}_{-i}(t)$, we have for $t\geq 1$:
$$
    \tilde{\pi}^{k}_{i}(t)=
\left\{
  \begin{array}{ll}
    1, & \hbox{if $m_{i}(t)$$<$$\mu$;} \\\\
   \pi^k_{i}(t), & \hbox{if $mod(t,2)=0;$} \\\\
     \pi^k_{i}(t)-d(1-2\epsilon)^{t-1}. & \hbox{if $mod(t,2)=1.$}
  \end{array}
\right.
$$
$$
    \tilde{\pi}^{k}_{-i}(t)=
\left\{
  \begin{array}{ll}
    1, & \hbox{if $m_{i}(t)$$<$$\mu$;} \\\\
   \pi^k_{-i}(t)-d(1-2\epsilon)^{t-1}, & \hbox{if $mod(t,2)=0;$} \\\\
     \pi^k_{-i}(t). & \hbox{if $mod(t,2)=1.$}
  \end{array}
\right.
$$
 Now, to accomplish the proof we need to define the associated expected utilities for each stage provided by $\pi_i(t)$ and $\tilde{\pi}_i(t)$, which are denoted by $u_{i,t}^{\star}$ and $\tilde{u}_{i,t}$, respectively.

\begin{equation}
u_{i,t}^{\star}= \ds{\sum_{a_0,a_1,a_2}} P_t(a_0,a_1,a_2) u_i(a_{0},a_{1},a_{2}),
\end{equation}
where: $P_t$ is the joint probability distribution, and $u_i$ the instantaneous utility (1). Denote by $u_{i}^{k}=u_i(a_{0},a_{1}^{\star},a_{2}^{\star})$, $u_{i}^{1}=u_i(a_{0},a_{1}^{\star},a^{\min})$, $u_{i}^{2}=u_{i}(a_{0},a^{\min},a_2^{\star})$, and
$u_{i}^{\min}=u_i(a_{0},a^{\min},a^{\min})$. $a_i^{\star}=f_i^{\star}(s_i)$, and $a^{\min}=(P_{\min},P_{\min})$. By means of these notations, we obtain:
\begin{multline}
u^{\star}_{i,t}=\ds{\sum_{a_0}}\rho(a_{0})[\pi_i^{k}(t)\pi_{-i}^{k}(t)u_i^{k}+\pi_i^{k}(t)(1-\pi_{-i}^{k}(t))u_i^{1}+\\(1-\pi_i^{k}(t))\pi_{-i}^{k}(t)u_i^{2}+(1-\pi_i^{k}(t))(1-\pi_{-i}^{k}(t))u_i^{\min}].
\label{eq91}
\end{multline}
We now define the expected utility of the deviation for each stage, denoted $\tilde{u}_{i,t}$.
\begin{multline}
\tilde{u}_{i,t}=\ds{\sum_{a_0}}\rho(a_{0})[\tilde{\pi}_i^{k}(t)\tilde{\pi}_{-i}^{k}(t)u_i^{k}+\tilde{\pi}_i^{k}(t)(1-\tilde{\pi}_{-i}^{k}(t))u_i^{1}+\\(1-\tilde{\pi}_i^{k}(t))\tilde{\pi}_{-i}^{k}(t)u_i^{2}+(1-\tilde{\pi}_i^{k}(t))(1-\tilde{\pi}_{-i}^{k}(t))u_i^{\min}].
\label{eq91}
\end{multline}
As the deviation distribution $\tilde{\pi}_i$ depends on the distribution provided by the proposed strategy $\sigma^{\star}_i$, $\pi_i$, one can also define $\tilde{u}_{i,t}$ as a function of $u^{\star}_{i,t}$, by using the definitions of $\pi_i(t)$ and
$\tilde{\pi}_i(t)$. Hence, we have the following result:
$$
    \tilde{u}_{i,t}=
\left\{
  \begin{array}{ll}
    u^{\star}_{i,t}, & \hbox{if $m_{i}(t)$$<$$\mu$;} \\\\
   u^{\star}_{i,t}-(1-2\epsilon)^{t-1}d\ds{\sum_{a_0}}\rho(a_{0})\breve{U}^1(t)&\hbox{if $mod(t,2)=0$;} \\\\

     u^{\star}_{i,t}-(1-2\epsilon)^{t-1}d\ds{\sum_{a_0}}\rho(a_{0})\breve{U}^2(t) &\hbox{if $mod(t,2)=1$,}
  \end{array}
\right.
$$

where: $\breve{U}^1(t)=\pi_{i}^k(t)(u_{i}^{k}-u_{i}^{1}+u_{i}^{\min}-u_{i}^{2})+ u_{i}^{2}-u_{i}^{\min}$, and $\breve{U}^2(t)=\pi_{-i}^k(t)(u_{i}^{k}-u_{i}^{1}+u_{i}^{\min}-u_{i}^{2})+ u_{i}^{1}-u_{i}^{\min}.$
Thus, the deviation utility of Node $i$ in the repeated game $\mathcal{\bar{G}}$ is:

$${U}_i(\tilde{\sigma}_i,{\sigma}_{-i}^{\star})=u_{i,0}^{\star}+\sum _{t=1}^{z}\delta^{t}\tilde{u}_{i,t}+\sum _{t=z+1}^{\infty}\delta^{t}u_{i,t}^{\star},$$
 where: $z$ is the number of stages until the condition $m_i<\mu$ is satisfied. The unilaterally deviation from the proposed strategy $\sigma^{\star}_i$ is not profitable if:
\begin{equation}\label{equilib}
    {U}_i(\tilde{\sigma}_i,{\sigma}_{-i}^{\star})\leq {U}_i({\sigma}_{i}^{\star},{\sigma}_{-i}^{\star}).
\end{equation}
The equilibrium condition could be determined using relation (\ref{equilib}). It is defined as follows:

$$ u_{i,0}^{\star}+\sum _{t=1}^{z}\delta^{t}\tilde{u}_{i,t}+\sum _{t=z+1}^{\infty}\delta^{t}u_{i,t}^{\star}\leq \sum _{t=0}^{\infty}\delta^{t}u_{i,t}^{\star}.$$

By substituting $\tilde{u}_{i,t}$ by its value, the equilibrium condition writes as:
\begin{multline}\label{append2}
\sum _{t=0}^{t=\frac{z}{2}-1}\delta^{2t}((1-2\epsilon)^{2t}d\ds{\sum_{a_0}}\rho(a_{0})\breve{U}^1(2t+1))\\
+\delta\sum _{t=0}^{t=\frac{z}{2}-1}\delta^{2t}((1-2\epsilon)^{2t+1}d\ds{\sum_{a_0}}\rho(a_{0})\breve{U}^2(2t+2))\geq 0.
\end{multline}

We have, $\breve{U}^1(2t+1)=\pi_{i}^k(2t+1)(u_{i}^{k}-u_{i}^{1}+u_{i}^{\min}-u_{i}^{2})+ u_{i}^{2}-u_{i}^{\min}$, and $\breve{U}^2(2t+2)=\pi_{-i}^k(2t+2)(u_{i}^{k}-u_{i}^{1}+u_{i}^{\min}-u_{i}^{2})+ u_{i}^{1}-u_{i}^{\min}$. We provide results for
$\pi_{i}^k(2t+1)=\pi_{-i}^k(2t+2)=1,$ which implies that relation (\ref{append2}) is satisfied for each $\pi_{i}^k(2t+1)$ and $\pi_{-i}^k(2t+2)$. With this assumption, the relation (\ref{append2}) becomes:

$$
\ds{\sum_{a_0}}\rho(a_{0})(u_{i}^{k}-u_{i}^{1})\sum _{t=0}^{t=\frac{z}{2}-1}\delta^{2t}(1-2\epsilon)^{2t}$$$$
+\ds{\sum_{a_0}}\rho(a_{0})(u_{i}^{k}-u_{i}^{2})\delta\sum _{t=0}^{t=\frac{z}{2}-1}\delta^{2t}(1-2\epsilon)^{2t+1}\geq0.
$$
This is satisfied if and only if:
\begin{equation}\label{condi}
\delta\geq\frac{\ds{\sum_{a_0}}\rho(a_{0})(u_{i}^{1}-u_{i}^{k})\sum _{t=0}^{t=\frac{z}{2}-1}\delta^{2t}(1-2\epsilon)^{2t}}{\ds{\sum_{a_0}}\rho(a_{0})(u_{i}^{k}-u_{i}^{2})(1-2\epsilon)\sum _{t=0}^{t=\frac{z}{2}-1}\delta^{2t}(1-2\epsilon)^{2t}}.
\end{equation}
The equilibrium condition is thus:
\begin{equation}\label{append21}
\delta\geq\max\left\{\frac{c_i}{(1-2\epsilon)r_i},0\right\},
\end{equation}

where: $c_i=\ds{\sum_{a_0}}\rho(a_{0})(u_{i}^{1}-u_{i}^{k})$, $r_i=\ds{\sum_{a_0}}\rho(a_{0})(u_{i}^{k}-u_{i}^{2})$, $u_{i}^{1}=u_i(a_{0},a_{1}^{\star},a^{\min})$, $u_{i}^{k}=u_i(a_{0},a_{1}^{\star},a_{2}^{\star})$, $u_{i}^{2}=u_{i}(a_{0},a^{\min},a_2^{\star})$, and $a_i^{\star}=f_i^{\star}(s_i)$.

Thus, the strategy profile  $(\sigma^{\star}_i,\sigma^{\star}_{-i})$  is a subgame perfect equilibrium if and only if:
\begin{equation}\label{append3}
\delta\geq\max\left\{\frac{c_i}{(1-2\epsilon)r_i},\frac{c_{-i}}{(1-2\epsilon)r_{-i}},0\right\}.
\end{equation}

\end{proof}

\end{document}